\documentclass[preprint,review,12pt]{elsarticle}
\biboptions{sort&compress}
\usepackage{graphicx}
\usepackage{subcaption}
\usepackage{float}
\usepackage{amssymb}
\usepackage{amsmath}

\usepackage{lineno}
\usepackage{booktabs}
\usepackage{verbatim}
\usepackage{adjustbox}
\usepackage{url}

\usepackage{xcolor}
\usepackage{caption}
\usepackage{enumitem}
\usepackage{hyperref}
\hypersetup{
    colorlinks=true,
    linkcolor=blue,
    filecolor=magenta,      
    urlcolor=cyan,
}
\journal{Powder Technology}
\begin{document}
\begin{frontmatter}
%

\title{Granular mixing and flow dynamics in horizontal stirred bed reactors}
%
%
\author{Sahar Pourandi\corref{cor1}}
\ead{s.pourandi@utwente.nl}

\author{Igor Ostanin}
\ead{i.ostanin@utwente.nl}
\author{Thomas Weinhart}
\ead{t.weinhart@utwente.nl}

\cortext[cor1]{Corresponding author.}

\address{University of Twente, Drienerlolaan 5, Enschede, The Netherlands}

\begin{abstract}
Horizontal stirred bed reactors (HSBRs) are widely used in gas--phase polyolefin production, where efficient solids mixing and controlled residence time distributions are essential for product quality and operational stability. Despite their industrial relevance, the influence of operating conditions on granular flow dynamics and mixing efficiency in HSBRs remains insufficiently understood. In this study, Discrete Element Method (DEM) simulations are used to investigate the effects of rotation speed and fill level on particle motion, mixing behaviour, and axial transport in a lab--scale HSBR.
An industrial--grade polypropylene powder is modelled using calibrated contact parameters. Mixing is quantified using the Lacey index in both axial (z) and cross--sectional (xy) directions. Particle circulation is characterised through a DEM--based cycle--time analysis and independently verified using a coarse--grained angular velocity field. Axial dispersion coefficients are obtained from particle trajectories using both an Einstein--type formulation and a cycle--based approach, and are further validated with a diffusion model predicting the evolution of the axial Lacey index.
Results show that axial mixing strongly depends on both rotation speed and fill level: increasing rotation speed accelerates axial homogenization, whereas higher fill levels lead to slower axial mixing. Cross--sectional mixing is primarily sensitive to rotation speed, with fill--level effects diminishing at higher speeds. Cycle time decreases with increasing rotation speed and fill level, indicating enhanced solids circulation. Axial dispersion increases with rotation speed but decreases with fill level, with the different evaluation methods yielding consistent results. These findings reveal a trade--off between axial mixing, circulation, and dispersion, highlighting the need to balance operating conditions when optimising HSBR performance and demonstrating the capability of DEM to support HSBR optimisation.
\end{abstract}


\end{frontmatter}

%
\section{Introduction}
Horizontal Stirred Bed Reactors (HSBRs) are widely used in industrial gas--phase polyolefin production, particularly for polypropylene, one of the two major commercial polyolefins alongside polyethylene, where efficient solids mixing and controllable residence--time distributions are essential for product quality and stable operation~\cite{shepard1976divided,caracotsios1992theoretical,soares2013polyolefin, POURANDI2026122404}. HSBRs originated from the early Amoco horizontal reactor concept~\cite{shepard1976divided} and have evolved into advanced industrial systems (e.g., Innovene and Horizon) employing segmented agitation to sustain continuous production at large scale~\cite{caracotsios1992theoretical,gorbach2000dynamics,soares2013polyolefin}. A key characteristic of HSBRs is that they combine plug--flow--like axial transport with strong internal mixing, leading to relatively narrow residence time distributions (RTDs) and improved control of polymer property distributions~\cite{zacca1996reactor,dittrich2007residence,tian2012modeling}.

From an operational perspective, HSBR performance is governed by the coupled dynamics of (i) mixing within cross--sections, (ii) solids circulation induced by the agitator, and (iii) axial transport and back--mixing along the reactor length. These mechanisms directly affect RTD shape and, consequently, polymer property uniformity and process stability~\cite{zacca1996reactor,dittrich2007residence,tian2012modeling}. Among the controllable operating parameters, impeller rotation speed and fill level are particularly important because they determine the balance between shear--driven agitation, bulk circulation, and axial migration of particles~\cite{gorbach2000dynamics,dittrich2007residence,van2024particle}.

Recent experimental studies using single--photon emission radioactive particle tracking (RPT) and X--ray imaging by van der Sande et al.~\cite{van2024particle,van2024flow} have provided direct evidence that particle motion and solids distribution in HSBRs are highly sensitive to operating conditions, including rotation speed and fill level. These measurements highlight how changes in operating conditions can alter circulation patterns and the degree of solids redistribution, potentially leading to non--uniform flow structures or stagnant regions when circulation is insufficient. Such observations motivate quantitative, systematic analysis of mixing and axial transport as functions of operating parameters.

Modelling efforts---including RTD analyses and reactor--scale frameworks for gas--phase polymerization---have further demonstrated that axial back--mixing, circulation, and local granular flow behaviour influence reactor performance and product distributions~\cite{caracotsios1992theoretical,gorbach2000dynamics,zacca1996reactor,dittrich2007residence,tian2012modeling}. However, establishing clear, parameter--resolved relationships between rotation speed and fill level on the one hand, and mixing efficiency, circulation, and axial transport on the other, remains challenging, partly because obtaining detailed internal flow information experimentally is difficult~\cite{van2024particle,van2024flow}.

Despite these advances, a detailed and systematic understanding of how rotation speed and fill level jointly influence mixing efficiency, particle circulation, and axial transport in HSBRs is still limited. This work addresses this gap by developing a Discrete Element Method (DEM) model that reproduces and explains the experimentally observed trends of van der Sande et al.~\cite{van2024particle}. Specifically, we investigate how rotation speed and fill level affect mixing efficiency, particle circulation, and axial transport in a lab--scale HSBR. By combining DEM simulations with multiple independent mixing and transport metrics, and by validating these against experimental observations, this work provides mechanistic insight into the coupled dynamics governing HSBR performance.

The paper is structured as follows. Section \ref{sec:methods} describes the methodologies used, including the DEM model, simulation procedure, and mixing and flowability analysis tools. Afterward, Section \ref{sec:results} presents the obtained results, measuring the Lacey index, cycle time, and axial and radial dispersion coefficients. Furthermore, we study the effect of operating parameters on mixing quality and particle motion. Finally, Section \ref{sec:conclusions} summarizes the main points of this work.
\section{Methods} \label{sec:methods}
\subsection{Particle characterization}

To model industrial conditions, we selected an industrial--grade polypropylene reactor powder as the granular material. The particle size distribution and aerated bulk density were characterized experimentally, as described in \cite{pourandi2024mathematical}. The particle size distribution (PSD), measured using laser diffraction, had a median diameter of approximately 904.6 µm, with the full distribution presented in Table \ref{tab:1}. The aerated bulk density was found to be $368\, \mathrm{kg/m^3}$, while the elastic modulus, shear modulus, and coefficient of restitution used in simulations were based on material databases, shown in Table \ref{tab:2}. Detailed procedures and characterization data are available in ref~\cite{pourandi2024mathematical}. 

\begin{table}[H]
\centering
\caption{Particle size distribution.}
\label{tab:1}
\begin{tabular}{ll}
\hline
\textbf{Particle diameter ($\mu$m)} & \textbf{Cumulative volume (\%)} \\
\hline
506.0 & 10 \\
712.7 & 25  \\
904.6 & 50 \\
1123 & 75 \\
1345 & 90 \\
\hline
\end{tabular}
\end{table}

\begin{table}[H]
\centering
\caption{Material properties of the polypropylene particles.}
\label{tab:2}
\begin{tabular}{ll}
\hline
\textbf{Parameter} & \textbf{Value}  \\
\hline
Elastic modulus $E$ & 1.325 $\times$ 10$^{9}$\,Pa\\
Shear modulus $G$ & 4 $\times$ 10$^{8}$\,Pa\\
Coefficient of restitution $e$ & 0.5\\
Aerated (loose) bulk density $\rho$ & 368\,kg/m$^{3}$ \\
\hline
\end{tabular}
\end{table}

\subsection{Simulation}

The Discrete Element Method (DEM) was used to simulate powder mixing in the lab--scale horizontal stirred bed reactor (HSBR). DEM allows for the accurate modeling of granular material behavior by representing particles as discrete entities and simulating their interactions \cite{cundall1979discrete}. The simulations were performed using the MercuryDPM software package \cite{weinhart2020fast, thornton2023recent}.

A combined Hertz--Mindlin model \cite{di2004comparison} and rolling friction model \cite{luding2008cohesive} were utilized for the DEM simulations of the HSBR, as previously described in \cite{pourandi2024mathematical}. To maintain consistent units, the rolling stiffness was defined as equal to the tangential stiffness used in the Mindlin model \cite{luding2008cohesive}.

To reduce computational cost while preserving the overall granular dynamics, the particle size in the HSBR simulations was scaled by a factor of 3, compared to a scaling factor of 5 used in \cite{pourandi2024mathematical}. This scaling approach accelerates simulations by reducing the total number of particles while maintaining similar bulk flow behaviour. As scaling influences contact interactions, the sliding ($\mu_s$) and rolling ($\mu_r$) friction coefficients were recalibrated under the present scaling conditions.

The calibration procedure described in \cite{pourandi2024mathematical} was followed, in which the friction coefficients are adjusted to ensure that the simulated angle of repose (AoR) matches experimental measurements. Based on this procedure, the values $\mu_s = 0.8$ and $\mu_r = 0.4$ were adopted for the simulations in this study.

\subsubsection{Simulation procedure}

\begin{figure}[H]
    \centering
    \renewcommand{\arraystretch}{1.2} 
    \begin{tabular}{c@{\hspace{2pt}}c@{\hspace{2pt}}c@{\hspace{0pt}}c}
        \raisebox{-.5\height}{\includegraphics[width=0.1\textwidth]{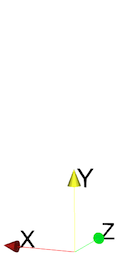}} &
        \raisebox{-.5\height}{\includegraphics[width=0.29\textwidth]{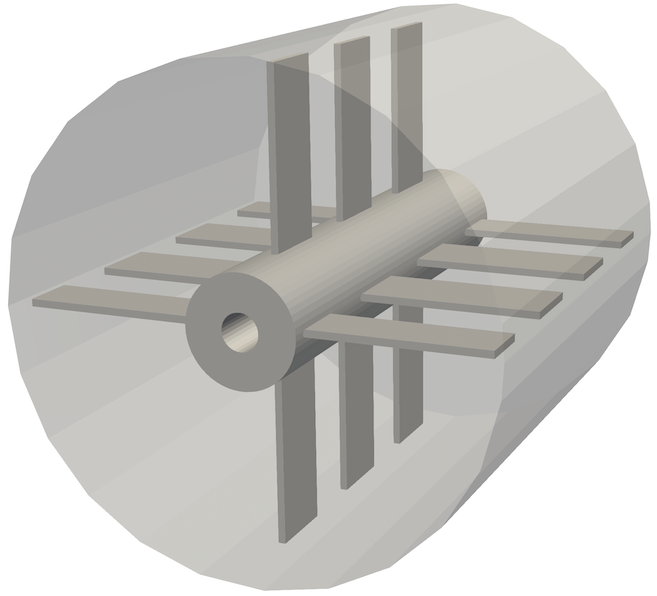}} &
        \raisebox{-.5\height}{\includegraphics[width=0.29\textwidth]{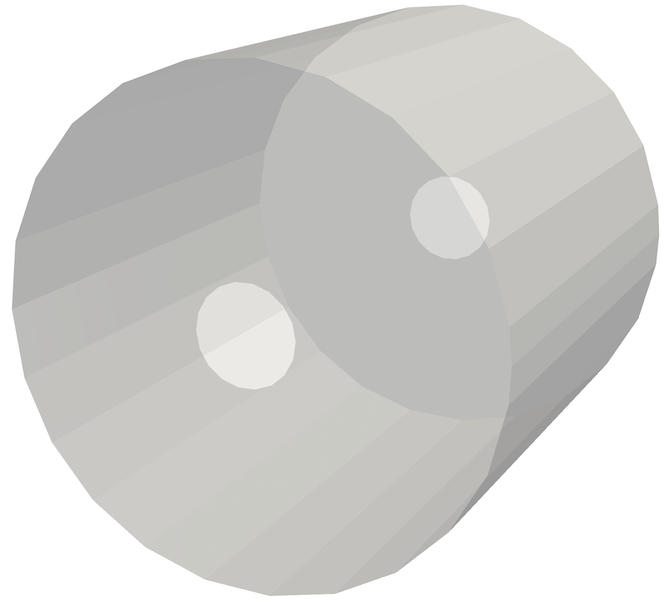}} &
        \raisebox{-.5\height}{\includegraphics[width=0.29\textwidth,clip,trim=17 0 0 0]{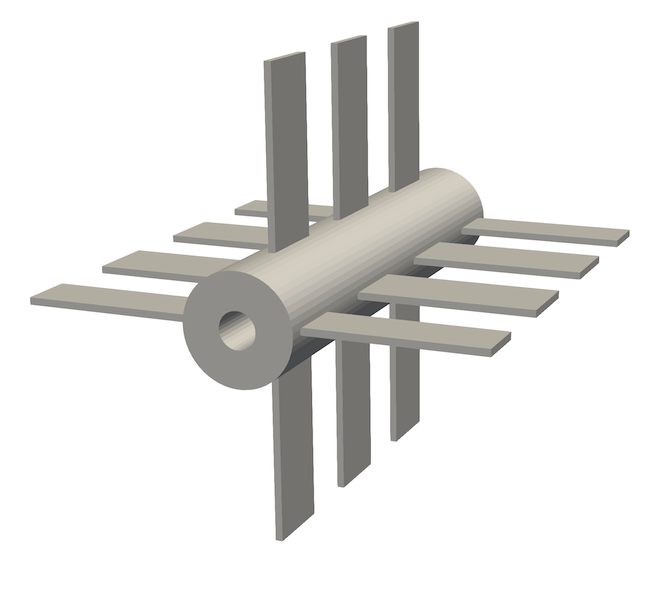}} \\
    \end{tabular}
    \caption{Lab--scale HSBR geometry. The column is fixed, and the agitator rotates.}
    \label{fig:HSBR-LabScale}
\end{figure}

The DEM simulations were conducted to evaluate the impact of fill level and rotation speed on particle dynamics and mixing efficiency in a lab--scale horizontal stirred bed reactor (HSBR). The simulated geometry and operating conditions were chosen to match the experimental setup reported by van der Sande et al.~\cite{van2024particle}, enabling direct comparison between simulation results and experimental measurements. The cylindrical reactor had an inner diameter of 134\,mm and a length of 150\,mm, with an agitator shaft featuring seven blade positions, each fitted with two blades positioned 90° apart. The inner blades were 20\,mm wide, while the end blades were 15\,mm wide. A schematic of the simulated HSBR is presented in Figure~\ref{fig:HSBR-LabScale}.

The interaction between particles and the reactor walls was modeled using the same contact properties as those applied for particle--particle interactions.

To investigate the influence of reactor parameters, simulations were conducted for combinations of four fill levels and six rotation speeds, as shown in Table \ref{tab:parameters}. The density in the simulations was set to ensure that the simulated bulk density matched the experimentally determined bulk density in \cite{pourandi2024mathematical}.

\begin{table}[H]
\centering
\caption{Fill levels and rotation speeds used in simulations.}
\label{tab:parameters}
\begin{tabular}{ll}
\hline
\textbf{Fill Level (\%)} & \textbf{Rotation Speed (rpm)} \\ \hline
40 & 10 \\ \hline
50 & 20 \\ \hline
60 & 30 \\ \hline
70 & 40 \\ \hline
   & 50 \\ \hline
   & 60 \\ \hline
\end{tabular}
\end{table}

Spherical particles were used in the DEM simulations, consistent with previous work \cite{pourandi2024mathematical}. This simplification was used to reduce computational complexity while preserving the essential particle dynamics. The shape effects of non--spherical particles were accounted for by adjusting the rolling friction coefficient (\(\mu_r\)) to higher values based on the relationship between particle shape and rolling resistance \cite{wensrich2012rolling, roessler2018scaling}. 

The simulation time step ($\Delta t$) was set to 20\% of the Rayleigh time step ($\Delta t_{\mathrm{Rayleigh}}$) to ensure numerical stability \cite{thakur2014micromechanical, huang2014time}. The definition and calculation procedure of $\Delta t_{\mathrm{Rayleigh}}$ are provided in Reference~\cite{pourandi2024mathematical}.

To further reduce computational time, the particle stiffness was decreased by a factor of $10^3$, following the guidelines of \cite{yan2015discrete} and \cite{lommen2014speedup}, without any effect on numerical stability or particle flow.

Simulations were conducted for all combinations of fill levels and rotation speeds, yielding a detailed dataset for analyzing the effects of these parameters on HSBR performance.

\subsection{Mixing and flowability analysis tools}\label{Method-Mixing and flowability analysis tools}

\subsubsection{Lacey index}

The Lacey index, first introduced by Lacey in 1954 \cite{lacey1954developments} and later refined in 1997 \cite{lacey1997mixing}, is one of the most widely used statistical indicators for quantifying the degree of homogeneity in particulate and powder mixing processes \cite{fuvesi2025mixing}. In Lacey’s formulation, the degree of mixing is defined as:

\begin{equation}
M = \frac{\sigma^2 - \sigma_0^2}{\sigma_r^2 - \sigma_0^2}
\label{eq:Lacey}
\end{equation}

\noindent where $\sigma^2$, $\sigma_0^2$, and $\sigma_r^2$ are the sample variances of the actual, fully segregated, and perfectly random states, respectively \cite{brandao2020experimental, fuvesi2025mixing}. This normalization ensures that $M$ is dimensionless and independent of sample size, enabling consistent comparisons across different systems and operating conditions.

The Lacey index provides a normalized measure of how closely a system approaches a random distribution relative to complete segregation. A value of $M = 0$ represents a fully segregated state, while $M = 1$ corresponds to a perfectly random (ideally mixed) system. 

The variance of the actual mixture, $\sigma^2$, is calculated from the composition of multiple spatial samples taken from the system:

\begin{equation}
\sigma^2 = \frac{1}{S - 1} \sum_{i=1}^{S} (f_{a,i} - f_a)^2
\end{equation}

\noindent where $S$ is the total number of samples, $f_{a,i}$ is the fraction of particles of type $a$ in the $i^{\text{th}}$ sample, and $f_a$ is the global fraction of type--$a$ particles in the system.

Following Lacey~\cite{lacey1997mixing}, the variance of a completely segregated state is defined as:

\begin{equation}
\sigma_0^2 = f_a (1 - f_a)
\end{equation}

\noindent since in the initial state, each sample contains only one type of particle. The variance for a perfectly random mixture---i.e., the statistical variance due to random fluctuations---is:

\begin{equation}
\sigma_r^2 = \frac{f_a (1 - f_a)}{n}
\end{equation}

\noindent where $n$ is the average number of particles per sample.

The Lacey index has become a standard metric in industrial mixing research due to its strong theoretical foundation and practical interpretability, and has been employed across a wide range of mixing systems in both batch and continuous processes. It has been applied in studies of rotating drums \cite{jiang2011enhancing, liu2013study, zhang2019numerical, brandao2020experimental, widhate2020mixing, patil2022dpm, van2024efficient, tang2024super, xu2024studying, song2025flow}, ribbon mixers \cite{chandratilleke2021study, jin2022investigation}, blade mixers \cite{chandratilleke2009effects, chandratilleke2012particle, chandratilleke2012study, halidan2014prediction, chandratilleke2014flow, chibwe2020particle}, as well as fluidized beds \cite{korkerd2024effect, godlieb2007characterizing} and tote or continuous blenders \cite{kamesh2022six}.

\begin{figure}[H]
  \centering
  \renewcommand{\arraystretch}{1.2}
  \begin{tabular}{c@{\hspace{12pt}}c}
    (a) & (b) \\
    \includegraphics[width=0.48\textwidth]{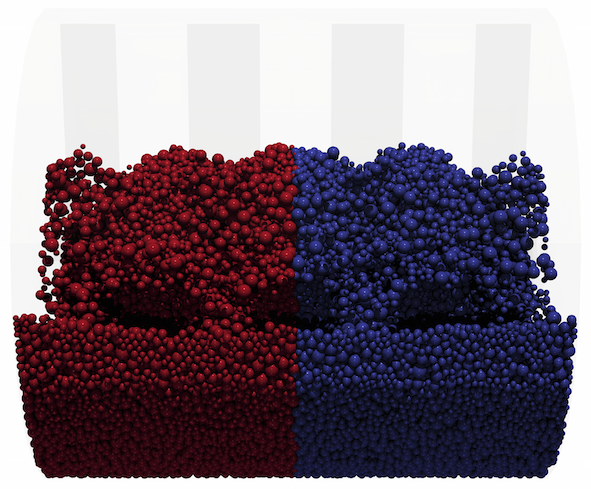} &
    \includegraphics[width=0.42\textwidth]{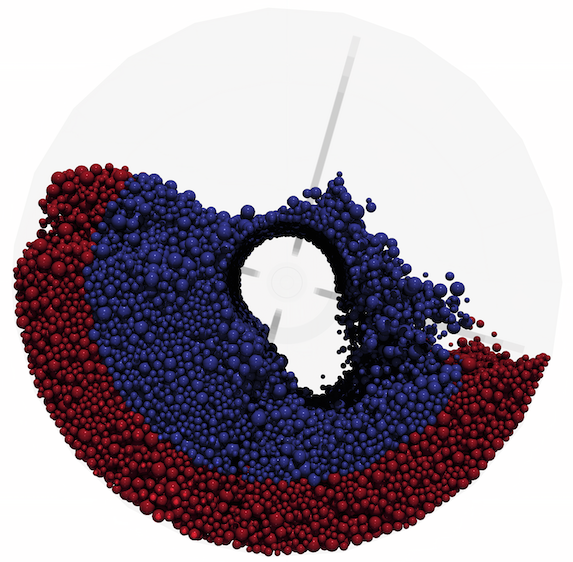} \\
  \end{tabular}
  \caption{Initial particle labeling used for mixing analysis: (a) axial (z--direction) labeling; (b) cross--sectional (xy--plane) labeling for the HSBR at 40\% fill level and 40\,rpm rotation speed.}
  \label{fig:Species}
\end{figure}

In this study, the Lacey index was employed to quantify both axial and cross--sectional mixing behavior in the horizontal stirred bed reactor (HSBR). A fully segregated initial condition was prepared by labelling particles according to their spatial position after the insertion phase, once all particles had been introduced into the reactor. To measure axial mixing, the reactor was divided into two equal halves: particles in the left half were labeled as red, and those in the right half as blue, as shown in Figure~\ref{fig:Species}(a). The domain was then segmented into ten axial bins using a mass--based sampling strategy such that each bin contains approximately equal total particle mass. The local mass fraction of red particles, $f_{a,i}$, was calculated in each bin over time and used in Equation~\ref{eq:Lacey} to compute the Lacey index $M(t)$. In this configuration, type--$a$ particles are defined as the red fraction, and since exactly half of the particles were initially assigned as red, the global composition was fixed at $f_a = 0.5$.

It is important to note that the labels ‘red’ and ‘blue’ are purely virtual identifiers based on initial particle position; both particle fractions have identical material properties. Therefore, given sufficient mixing time, these two labeled populations are expected to fully homogenize, providing a reliable basis for assessing mixing performance.

To measure cross--sectional mixing, particle labeling was defined radially, as illustrated in Figure~\ref{fig:Species}(b). Particles located closer to the reactor center were labeled as blue, while those in the outer radial region, near the vessel wall, were labeled as red. The cross--section was partitioned into $10 \times 10$ bins using a mass--based sampling strategy, and the red particle mass fraction was computed in each bin throughout the simulation to evaluate the Lacey index over time.

This sampling strategy partitions the domain into bins with approximately equal total particle mass, aiming to keep particle counts per bin reasonably balanced and to reduce statistical noise in variance estimation. The time evolution of $M(t)$ effectively captures the transition from a fully segregated state ($M = 0$) to an ideally mixed state ($M \approx 1$). By analyzing $M(t)$ under varying operational parameters such as impeller rotation speed and fill level, the efficiency of granular mixing can be quantitatively characterized. Owing to its normalization and theoretical rigor, the Lacey index continues to serve as a reliable and widely accepted measure of mixing quality in both experimental studies and DEM simulations of mechanically agitated systems, such as the HSBR examined here.

\subsubsection{Cycle time}

Cycle--time analysis has been employed as a practical tool to investigate particle circulation dynamics in mechanically agitated granular systems, particularly horizontal stirred bed reactors (HSBRs). Van~der~Sande et al.~\cite{van2024particle, van2024flow} introduced this concept experimentally to characterise solids circulation in HSBRs using both radioactive particle tracking and X--ray imaging. Their studies established cycle time as a useful indicator of solids circulation and flowability within the reactor. Similar circulation concepts have been discussed in earlier granular mixing research, where the cycle time represents the period required for a particle to complete one full revolution around the agitator or vessel circumference~\cite{norouzi2015insights}. A shorter cycle time means the solids recirculate more quickly, enabling more frequent particle exchange and typically enhancing radial and circumferential mixing. Conversely, longer cycle times are associated with slower circulation and the persistence of localized circulation loops or stagnant zones~\cite{norouzi2015insights,van2024particle,van2024flow}.

In continuous HSBRs, cycle time is also linked to macroscopic transport behaviour: rapid internal cycling promotes solids back--mixing and broadens the residence time distribution, whereas slower cycling tends to yield more plug--flow--like operation~\cite{zacca1996reactor}. Experimental measurements using radioactive particle tracking technique~\cite{van2024particle} and numerical studies employing the Discrete Element Method (DEM)~\cite{norouzi2015insights} have demonstrated that operating conditions such as rotational speed and fill level strongly influence circulation time and, consequently, flowability.

\begin{figure}[H]
    \centering
    \includegraphics[width=0.6\textwidth]{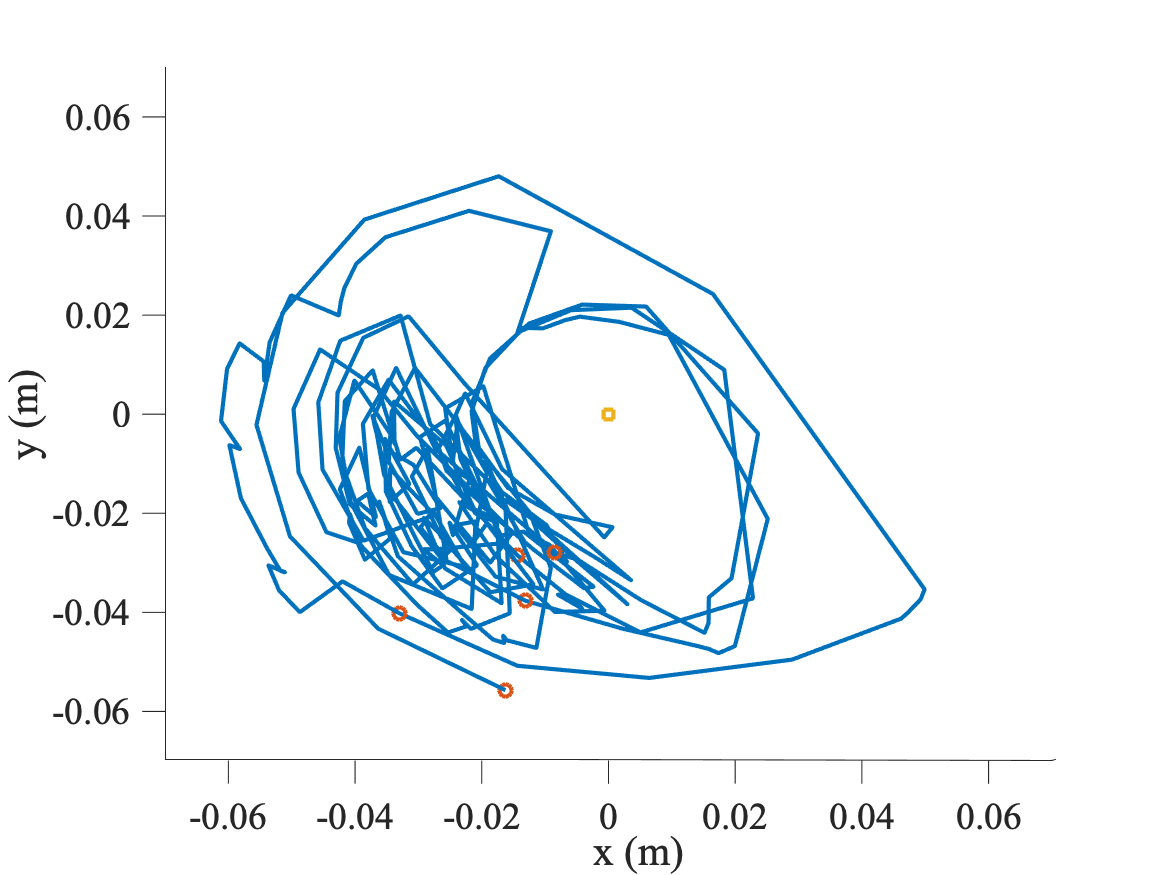}
    \caption{Sample particle trajectory in the xy--plane (blue), particle positions at $t^\mathrm{c}_{i,k}$ (red), and shaft center (yellow). Trajectories are linearly interpolated between simulation time steps.}
    \label{fig:ParticleTrajectory}
\end{figure}

In this study, we extend this concept in silico by implementing a DEM--based definition of cycle time to quantify angular transport across different HSBR operating scenarios. The cycle time $\Delta t^\mathrm{c}_{i,k}$ is defined as the time required for particle $i$ to complete its $k^\text{th}$ revolution around the central shaft, i.e.,

\begin{equation}
\Delta t^\mathrm{c}_{i,k} = t^\mathrm{c}_{i,k+1} - t^\mathrm{c}_{i,k}
\end{equation}

\noindent where $t^\mathrm{c}_{i,k}$ is the time at which particle $i$ completes its $k^\text{th}$ cycle. In this study, a full revolution is identified when the particle has sequentially traversed all four quadrants in the xy--plane. This criterion is based on angular displacement rather than returning to the exact same geometric location or angle. As a result, the angular positions of the particle at successive $t^\mathrm{c}_{i,k}$ may differ, as shown in Figure~\ref{fig:ParticleTrajectory}.

To avoid counting the transient period during which particles are still being inserted into the mixer, only revolutions after the insertion phase are considered in the analysis.

The mean cycle time, averaged over all particles and their valid revolutions, is computed as:

\begin{equation}
\overline{\Delta t^\mathrm{c}} = \frac{\sum_{i=1}^{N_\mathrm{p}} \sum_{k=1}^{N^\mathrm{c}_i} \Delta t^\mathrm{c}_{i,k}}{\sum_{i=1}^{N_\mathrm{p}} N^\mathrm{c}_i}
\label{eq:MeanCycleTime}
\end{equation}

\noindent Here, $N_\mathrm{p}$ is the total number of particles in the system, and $N^\mathrm{c}_i$ is the number of completed revolutions detected for particle $i$.

Figure~\ref{fig:ParticleTrajectory} illustrates the xy trajectory of a sample particle (blue line), with red circles marking the positions at each $t^\mathrm{c}_{i,k}$. The yellow circle marks the center of the shaft. Note that the particle positions at $t^\mathrm{c}_{i,k}$ are not located at the same angle or distance from the center. This reflects the fact that the revolution condition is based on passing through all four angular quarters, not exact spatial repetition. Additionally, the blue trajectory segments are formed via linear interpolation between simulation output time steps.

\paragraph{Cycle Time Verification via Coarse--Graining Analysis}
To verify the trajectory-based measurement of the mean cycle time, a continuum coarse--graining (CG) analysis was also implemented using the \texttt{fstatistics} tool of the MercuryCG framework~\cite{weinhart2012discrete}. The DEM particle data were converted into continuum fields by applying a Gaussian smoothing kernel with a bandwidth equal to one mean particle radius, yielding spatially coarse--grained fields for the solid volume fraction $\phi(x,y)$ and the velocity components $v_x(x,y)$ and $v_y(x,y)$ in the xy--plane.

The tangential velocity in the angular direction was computed from the coarse--grained fields as

\begin{equation}
v_\theta(x,y)
    = \frac{x\,v_y - y\,v_x}{\sqrt{x^2+y^2}},
\qquad
r = \sqrt{x^2+y^2},
\label{eq:vthetaCG_Method}
\end{equation}

\noindent and the corresponding angular velocity field followed directly as

\begin{equation}
\omega(x,y) = \frac{v_\theta(x,y)}{r}.
\label{eq:omegaCG_Method}
\end{equation}

The connection between this coarse--grained angular velocity field and the mean cycle time can be derived from the angular momentum balance. The angular mass flux through the cross--section $R^2$ is

\begin{equation}
\dot{m}
  = \frac{1}{2\pi} \int_{R^2} \frac{v_\theta}{r}\,\rho\, dA,
\label{eq:mdot_Method}
\end{equation}

\noindent where $\rho$ denotes the coarse--grained mass density. The total mass in the same cross--section is

\begin{equation}
m = \int_{R^2} \rho\, dA.
\label{eq:mCG_Method}
\end{equation}

\noindent These two quantities define the mass--weighted mean angular velocity,

\begin{equation}
\bar{\omega}
  = \frac{\displaystyle\int_{R^2} \omega\,\rho\, dA}
         {\displaystyle\int_{R^2} \rho\, dA}.
\label{eq:omegaMass_Method}
\end{equation}

Since the particle material density $\rho_p$ is constant, the coarse--grained mass density satisfies $\rho = \rho_p \phi$, and the mass--weighted average becomes equivalent to a volume--fraction--weighted average:

\begin{equation}
\bar{\omega}
  = \frac{\displaystyle\int_{R^2} \omega(x,y)\,\phi(x,y)\, dA}
           {\displaystyle\int_{R^2} \phi(x,y)\, dA}
  \approx \frac{\langle \phi\,\omega\rangle}{\langle \phi\rangle}.
\label{eq:omegaVF_Method}
\end{equation}

Finally, the mean cycle time associated with the coarse--grained angular velocity field follows from the ratio of the total mass to the angular mass flux:

\begin{equation}
\overline{\Delta t^{\mathrm{c}}}_{\mathrm{CG}}
    = \frac{m}{\dot{m}}
    = \frac{2\pi \displaystyle\int \rho\,dA}
           {\displaystyle\int \omega\,\rho\,dA}
    = \frac{2\pi}{\bar{\omega}},
\label{eq:tcCG_Method}
\end{equation}

\noindent which represents the time required for a mass--weighted average particle to complete one full revolution around the reactor shaft.

The coarse--grained estimate $\overline{\Delta t^{\mathrm c}}_{\mathrm{CG}}$ provides an independent prediction of the mean cycle time that can be directly compared with the trajectory--based value obtained from particle revolutions. In the Results section, the agreement between these two measurements is used to confirm that the DEM-based definition of cycle time reliably captures the angular circulation dynamics of the HSBR.

\subsubsection{Axial dispersion coefficient}

In addition to the Lacey index and cycle--time analysis, the axial dispersion coefficient provides a complementary measure of how efficiently particles are transported along the reactor length. Rather than describing the overall degree of mixing, axial dispersion quantifies the stochastic spreading of particle trajectories around the mean axial drift. This behaviour is directly linked to the solids residence time distribution and the extent of axial back-mixing in HSBRs, both of which influence process stability and product uniformity. Previous modelling and experimental studies have shown that axial transport strongly affects polymer properties, catalyst distribution, and thermal behaviour in gas--phase polyolefin reactors \cite{zacca1996reactor, dittrich2007residence, caracotsios1992theoretical, soares2013polyolefin}. In practice, an appropriate level of axial dispersion helps maintain smooth solids flow along the reactor, prevents local accumulation or maldistribution of catalyst, and supports the desired balance between plug-flow-like behaviour and controlled back-mixing \cite{tian2012modeling, gorbach2000dynamics}.

In the literature, two main approaches are typically distinguished for evaluating dispersion coefficients in particulate systems: a \emph{macro--scale} approach and a \emph{micro--scale} approach~\cite{yang2016influence,yang2018investigation}. In the macro approach, the dispersion coefficient is obtained by fitting the evolution of concentration profiles to a Fickian advection--diffusion model at the continuum scale. In the micro approach, the dispersion is computed directly from particle displacements, following Einstein’s original treatment of Brownian motion~\cite{einstein1905molekularkinetischen, einstein1956investigations} and its subsequent extensions to granular flows~\cite{luo2013particle,yang2016influence,yang2018investigation}. Since DEM provides full particle trajectories, the micro (Einstein--type) formulation is adopted here.

\paragraph{Time--Based (Einstein) Formulation}
Following this micro--scale approach, the instantaneous axial dispersion of particle $i$ over a sampling interval $[t_j,t_j+\Delta t]$ is defined as

\begin{equation}
D_{i,j}^{z}(\Delta t)
    = \frac{\left( \Delta z_{i,j} - \overline{\Delta z} \right)^{2}}
           {2\,\Delta t},
\label{eq:EinsteinDispersion_ij}
\end{equation}

\noindent where the axial displacement over one sampling interval is

\[
\Delta z_{i,j} = z_i(t_j+\Delta t) - z_i(t_j),
\]

\noindent where, assuming a statistically steady regime, $\overline{\Delta z}$ is the mean axial displacement, averaged over all particles and all sampling intervals,

\begin{equation}
\overline{\Delta z}
   = \frac{1}{N_t N_p}\sum_{j=1}^{N_t}\sum_{i=1}^{N_p} \Delta z_{i,j}.
\label{eq:MeanDz_Einstein}
\end{equation}

\noindent Here, $N_p$ is the number of particles and $N_t$ is the number of sampled time steps.

The axial dispersion coefficient of the system at time $t_j$ is then obtained by averaging over all particles,

\begin{equation}
D_j^{z}
   = \frac{1}{N_p}\sum_{i=1}^{N_p} D_{i,j}^{z},
\label{eq:EinsteinDispersion_j}
\end{equation}

\noindent and the time--averaged steady--state axial dispersion coefficient $D^{z}$ of the whole system, as in Yang et al.~\cite{yang2016influence}, is obtained by averaging $D_j^{z}$ over all sampling intervals,

\begin{equation}
D^{z}
  = \frac{1}{N_t}\sum_{j=1}^{N_t} D_j^{z}.
\label{eq:EinsteinDispersion_final}
\end{equation}

This time--based (Einstein) formulation is used as the primary definition of axial dispersion throughout this thesis. 

\paragraph{Cycle--Based Formulation}
An alternative dispersion formulation can be derived using particle displacements evaluated cycle--by--cycle, following the methodology of Ingram et al.~\cite{ingram2005axial}. For particle $i$, the axial displacement during its $k$-th cycle is

\begin{equation}
\Delta z_{i,k}
    = z_i\!\left(t^{c}_{i,k+1}\right)
    - z_i\!\left(t^\mathrm{c}_{i,k}\right),
\label{eq:dz_cycle}
\end{equation}

\noindent where $t^\mathrm{c}_{i,k}$ is the time at which the particle completes its $k^{\mathrm{th}}$ revolution around the mixer shaft.

The mean cycle--wise displacement is

\begin{equation}
\overline{\Delta z}
   = \frac{\displaystyle\sum_{i=1}^{N_p}\sum_{k=1}^{N_i^{c}} \Delta z_{i,k}}
           {\displaystyle\sum_{i=1}^{N_p} N_i^{c}},
\label{eq:dzbar_cycle}
\end{equation}

\noindent which approaches zero due to the geometric symmetry of the HSBR.

The cycle--based axial dispersion coefficient is then

\begin{equation}
D^{z}_{\mathrm{c}}
  = \frac{\displaystyle\sum_{i=1}^{N_p}\sum_{k=1}^{N_i^{c}}
        \left( \Delta z_{i,k} - \overline{\Delta z} \right)^{2}}
         {2 \displaystyle\sum_{i=1}^{N_p} N_i^{c}}.
\label{eq:Dz_cycle}
\end{equation}

This definition treats each particle cycle as an independent sampling interval.
Although both formulations are physically consistent, the time--based definition is used as the primary dispersion measure in the Results, while the cycle--based formulation serves as an independent verification.

\paragraph{Choice of Sampling Interval \boldmath$\Delta t$}
To determine an appropriate sampling interval, the dependence of the time--based dispersion coefficient $D^{z}$ on $\Delta t$ was evaluated over a wide range of $\Delta t$ values. The resulting curves are presented in \ref{app:DeltaTStudy}, Figure~\ref{fig:Dz_vs_deltat}. For very small $\Delta t$, the calculated dispersion coefficient decreases rapidly because particle motion remains strongly correlated, in agreement with the observations of Pallarès and Johnsson~\cite{pallares2006novel}. For intermediate $\Delta t$, a quasi--plateau region is observed in which the dispersion coefficient becomes only weakly dependent on $\Delta t$. For very large $\Delta t$, the curves corresponding to different operating conditions gradually approach each other, indicating a loss of resolution in the dispersion measurement, as also reported for experimental HSBRs by van~der~Sande et al.~\cite{van2024particle}.

In the present DEM simulations, a slight residual decrease of the dispersion coefficient with increasing $\Delta t$ is observed even within the quasi--plateau region. This behaviour is consistent with a sub--linear scaling of the mean-squared axial displacement, $\langle \Delta z^2 \rangle \propto \Delta t^{\alpha}$. The fitted exponent $\alpha < 1$ (\ref{app:DeltaTStudy}, Figure~\ref{fig:MSD_axial}) indicates sub-diffusive particle motion.

Based on the existence of a broad quasi-plateau region, and to retain sufficient temporal resolution while preserving sensitivity to operating conditions, a sampling interval of $\Delta t = 50\,\mathrm{s}$ was selected for all time--based dispersion calculations in this study. This choice follows the same physical criteria used in the experimental HSBR study of van~der~Sande et al.~\cite{van2024particle}, while accounting for the slightly slower decorrelation dynamics observed in the present DEM system with scaled particles.

\paragraph{Axial Dispersion Verification via a Diffusion Model}
To independently verify the axial dispersion coefficient obtained from the particle--based formulations, a one-dimensional diffusion model was solved in the axial direction. The solids transport was approximated by a Fickian diffusion equation for the local volume (or number) fraction of type--$a$ particles, $c(z,t)$,

\begin{equation}
  \frac{\partial c}{\partial t}
    = D^{z} \,\frac{\partial^2 c}{\partial z^2},
  \label{eq:AxialDiffusionPDE}
\end{equation}

\noindent where $z$ is the axial coordinate and $D^{z}$ is the constant axial dispersion coefficient to be tested. The domain length matched the DEM reactor geometry, $0 \le z \le L_z$.

The initial condition was chosen to mirror the axial labeling used for the DEM-based Lacey index. The reactor was split into two equal axial halves: the left half fully occupied by type--$a$ particles and the right half by type--$b$ particles,

\begin{equation}
  c(z,0) =
  \begin{cases}
    1, & 0 \le z < L_z/2, \\
    0, & L_z/2 \le z \le L_z,
  \end{cases}
  \label{eq:AxialDiffusionIC}
\end{equation}

\noindent corresponding to a fully segregated system with global composition $f_a = 0.5$.

Zero-flux (Neumann) boundary conditions were imposed at both ends of the reactor:

\begin{equation}
  \left.\frac{\partial c}{\partial z}\right|_{z=0}
    = 0, \qquad
  \left.\frac{\partial c}{\partial z}\right|_{z=L_z}
    = 0,
  \label{eq:AxialDiffusionBC}
\end{equation}

\noindent preventing any net axial flux of labeled particles. Equation~\eqref{eq:AxialDiffusionPDE} with the initial and boundary conditions above was solved numerically using the \texttt{pdepe} solver in MATLAB on a uniform axial grid.

To compare the diffusion-model predictions with the DEM results, the numerical solution $c(z,t)$ was post--processed to compute a model prediction of the axial Lacey index. The axial coordinate was divided into the same $S$ axial bins used in the DEM analysis, and the local fraction of type--$a$ particles in bin $i$ was taken as

\[
f_{a,i}(t) \approx c(z_i,t),
\]

\noindent where $z_i$ is the bin center. From these sampled concentrations, the Lacey index was computed using the definition introduced in Section~\ref{Method-Mixing and flowability analysis tools}. Because the diffusion equation produces a smooth continuum concentration field without sampling noise, the perfectly random--state variance $\sigma_r^2$ tends to zero, consistent with the continuum limit of the Lacey formulation.

This procedure provides a model prediction of the axial Lacey index $M_z(t)$ for any chosen value of the dispersion coefficient $D^{z}$. In the Results section, the diffusion--based prediction of $M_z(t)$ is directly compared with the corresponding $M_z(t)$ obtained from the DEM particle data. The agreement between these curves is used to assess the consistency of the independently computed dispersion coefficients (time--based (Einstein) and cycle--based) and to verify the selected value of $D^{z}$.
\section{Results and discussion}\label{sec:results}
\subsection{Mixing and flowability measurements}

\subsubsection{Lacey index}

\begin{figure}[H]
  \centering
  \renewcommand{\arraystretch}{1.2}
  \begin{tabular}{c@{\hspace{10pt}}c} 
    \includegraphics[width=0.47\textwidth]{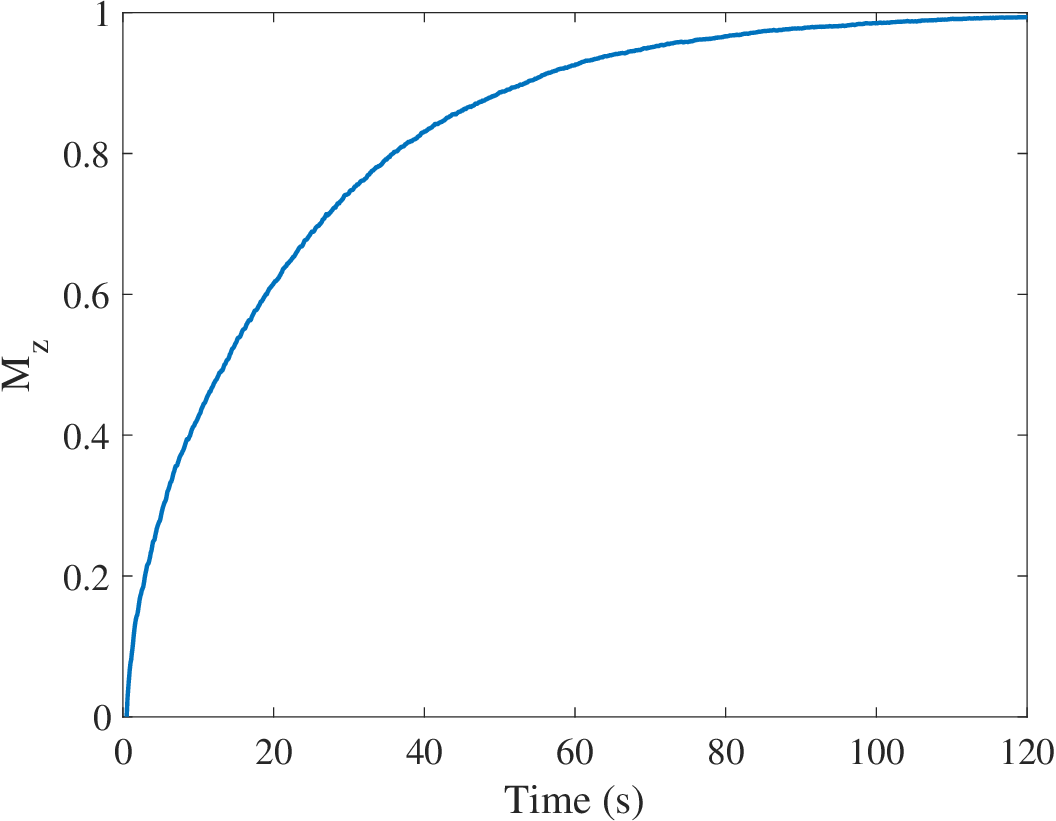} &
    \includegraphics[width=0.47\textwidth]{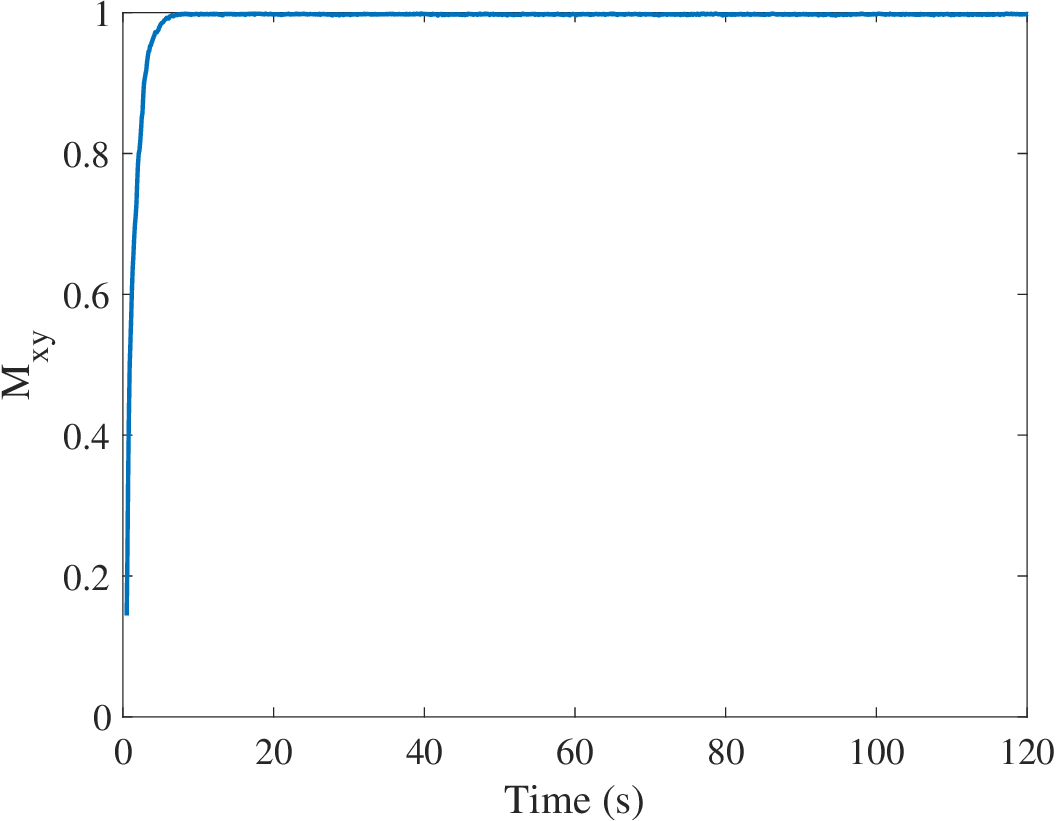} \\
  \end{tabular}
  \caption{Lacey index in the axial direction (left) and cross--sectional direction (right) for the HSBR at a fill level of 40\% and rotation speed of 40\,rpm.}
  \label{fig:M}
\end{figure}

As outlined in Section~\ref{Method-Mixing and flowability analysis tools}, the Lacey index was used to monitor the evolution of mixing over time in the horizontal stirred bed reactor (HSBR). Figure~\ref{fig:M} shows how the Lacey index evolved during mixing at 40\,rpm rotation speed with a 40\% fill level of polypropylene powder.

The plot on the left side of Figure~\ref{fig:M} shows the axial mixing behavior. The Lacey index gradually increases and approaches 0.99 after about 102 seconds (68 rotations), indicating that the system reaches a nearly homogeneous state in the axial direction within that time.

In contrast, the plot on the right shows much faster mixing in the cross--sectional (xy) plane. Here, the Lacey index reaches 0.99 in just 6 seconds (4 rotations), reflecting a rapid approach to uniformity.

This stark difference in mixing time reflects the dominant transport mechanisms in each direction. Cross--sectional mixing is primarily driven by convective motion from the impeller blades, supplemented by dispersion, leading to rapid homogenization. In contrast, axial mixing relies predominantly on slower dispersive transport. As a result, mixing in the transverse plane is significantly faster than in the axial direction.

\begin{figure}[H]
    \centering
    \renewcommand{\arraystretch}{1.2}
    \begin{tabular}{c@{\hspace{4pt}}c@{\hspace{4pt}}c@{\hspace{4pt}}c}
        \includegraphics[width=0.24\textwidth]{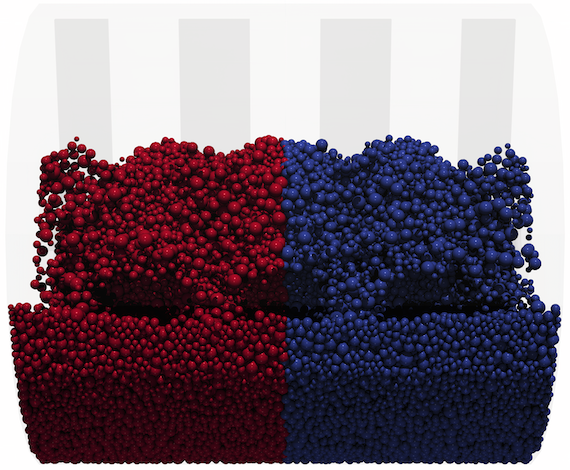} &
        \includegraphics[width=0.24\textwidth]{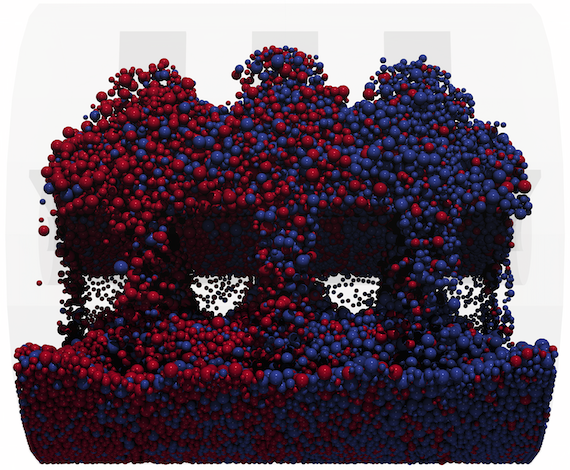} &
        \includegraphics[width=0.24\textwidth]{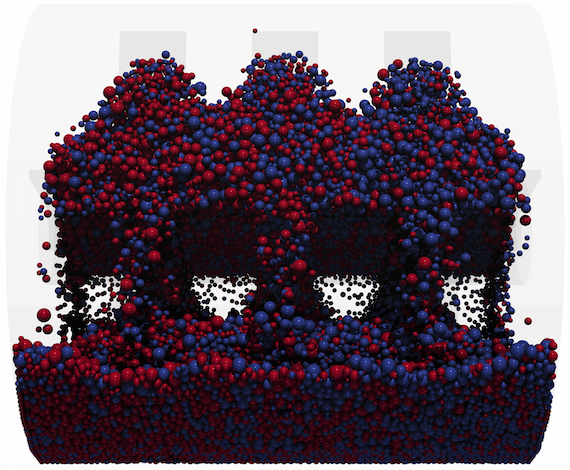} &
        \includegraphics[width=0.24\textwidth]{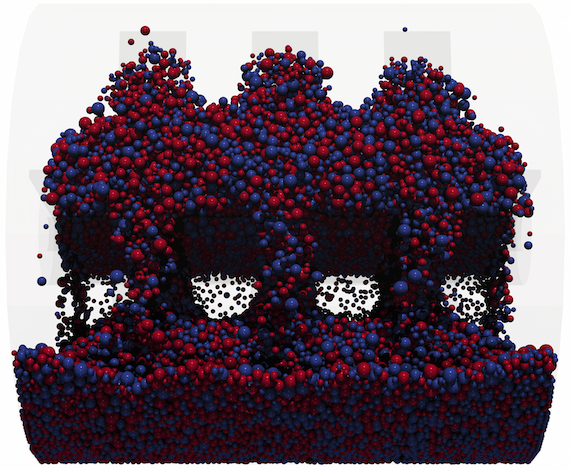} \\
        0.5 s & 30 s & 60 s & 102 s
    \end{tabular}
    \caption{Simulation snapshots of axial mixing evolution in the HSBR at a fill level of 40\% and rotation speed of 40\,rpm.}
    \label{fig:HSBR_AxialSpecies}
\end{figure}

\begin{figure}[H]
    \centering
    \renewcommand{\arraystretch}{1.2}
    \begin{tabular}{c@{\hspace{4pt}}c@{\hspace{4pt}}c@{\hspace{4pt}}c}
        \includegraphics[width=0.24\textwidth]{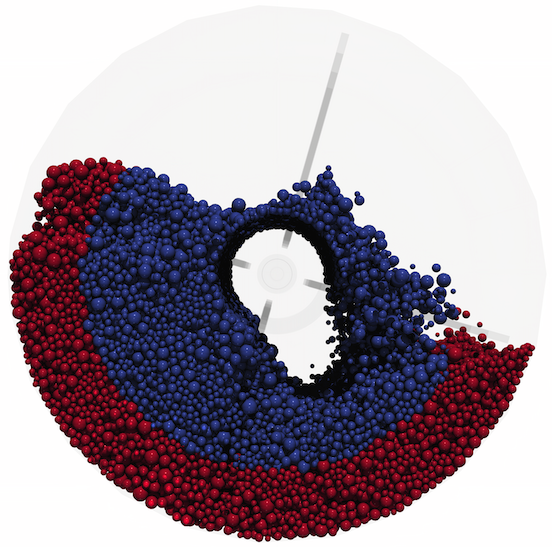} &
        \includegraphics[width=0.24\textwidth]{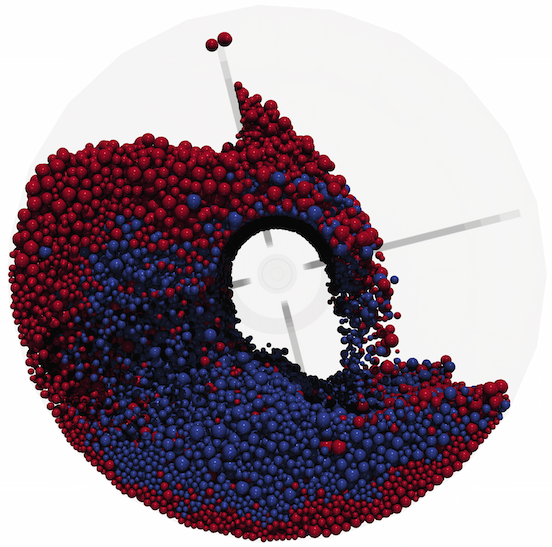} &
        \includegraphics[width=0.24\textwidth]{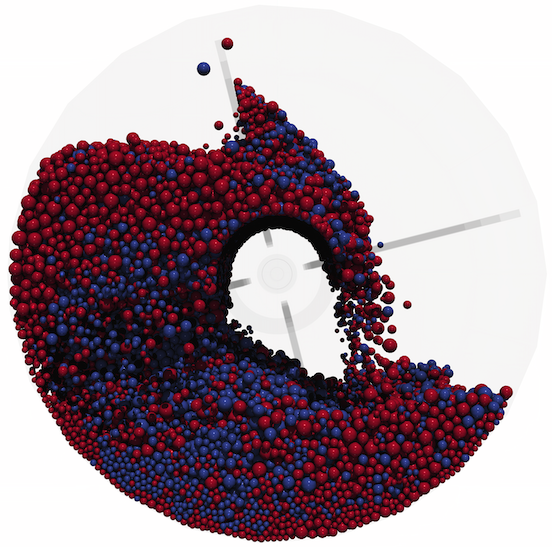} &
        \includegraphics[width=0.24\textwidth]{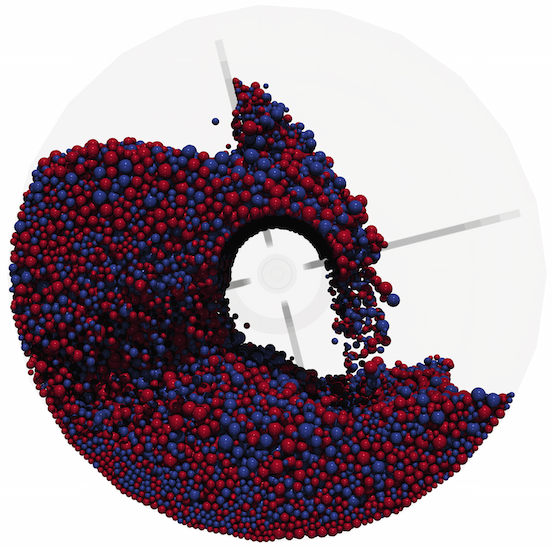} \\
        0.5 s & 1.5 s & 3 s & 6 s
    \end{tabular}
    \caption{Simulation snapshots of cross--sectional mixing evolution in the HSBR at a fill level of 40\% and rotation speed of 40\,rpm.}
    \label{fig:HSBR_RadialSpecies}
\end{figure}

To visually illustrate the progression of mixing, Figures~\ref{fig:HSBR_AxialSpecies} and~\ref{fig:HSBR_RadialSpecies} present simulation snapshots of particle distributions at selected time points in the axial and cross--sectional directions, respectively. These snapshots complement the Lacey index results by providing a visual representation of how the initially segregated red and blue particles become increasingly intermixed over time.

In the axial direction (Figure~\ref{fig:HSBR_AxialSpecies}), even after 30 seconds, large regions of red and blue particles remain clearly distinguishable, with only limited interpenetration between the two zones. By 60 seconds, some red particles have dispersed into the blue region and vice versa, and by 102 seconds, the particle distribution appears largely uniform. This gradual transition aligns with the slower rise and eventual plateau of the axial Lacey index, highlighting dispersive transport as the primary mixing mechanism along the reactor’s length.

In the cross--sectional plane (Figure~\ref{fig:HSBR_RadialSpecies}), mixing proceeds much more rapidly. At 1.5 seconds, partial intermixing is evident, but distinct red and blue clusters still exist. At 3 seconds, segregation is still visible, yet clear convective movement can be observed as particles are rotated and redistributed. By 6 seconds, the system achieves near-complete homogenization. These patterns align with the sharp rise in the radial Lacey index and reflect the combined influence of convective transport, driven by the rotating blades, and localized particle dispersion, which together promote rapid radial mixing.

\paragraph {Mixing Time}
To quantify the mixing behavior observed in both the axial (z) and cross--sectional (xy) directions, the time evolution of the Lacey index was fitted to an exponential model of the form \cite{liu2013effect,halidan2018mixing,herman2021effect}:

\begin{equation}
    M^{\text{fit}}(t) = 1 - \exp\left(-\frac{t - t_0}{t_\mathrm {m}}\right)
\end{equation}

\noindent Here, $t$ denotes time, $t_0$ is the offset time representing the delay before mixing begins, and $t_\mathrm {m}$ is the characteristic mixing time. Smaller values of $t_\mathrm {m}$ indicate faster mixing, while larger values reflect slower homogenization.

\begin{figure}[H]
  \centering
  \renewcommand{\arraystretch}{1.2}
  \begin{tabular}{c@{\hspace{10pt}}c}
    \includegraphics[width=0.47\textwidth]{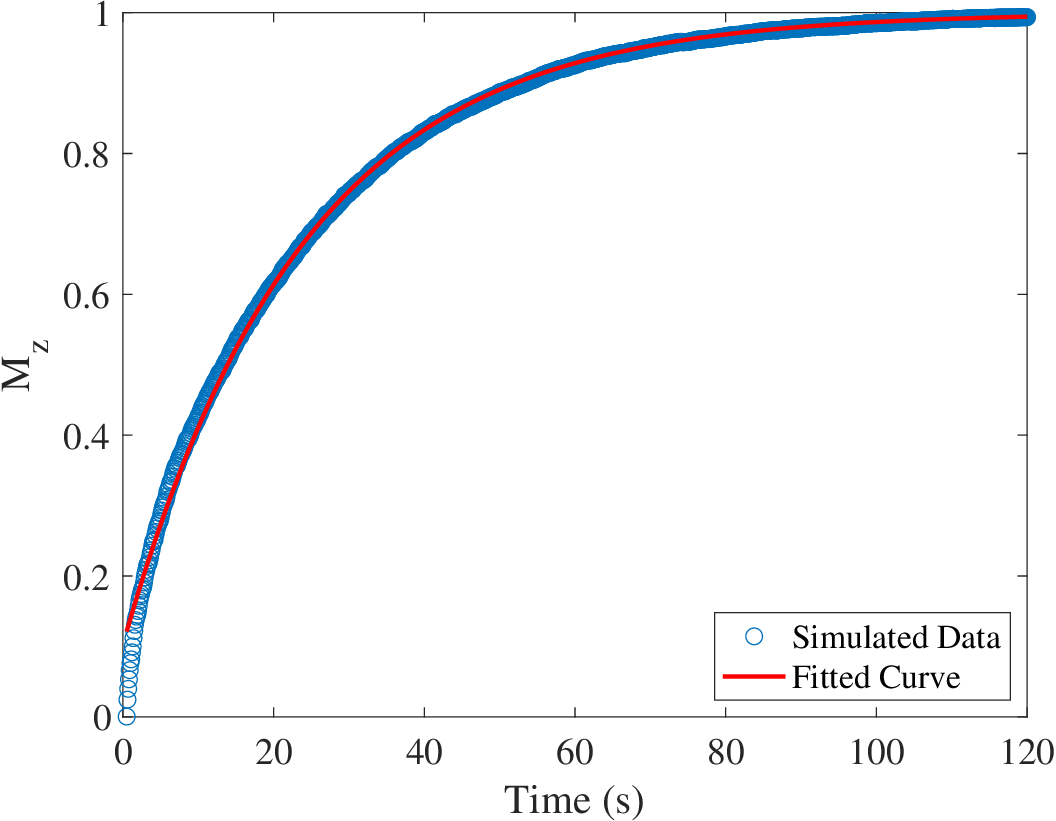} &
    \includegraphics[width=0.47\textwidth]{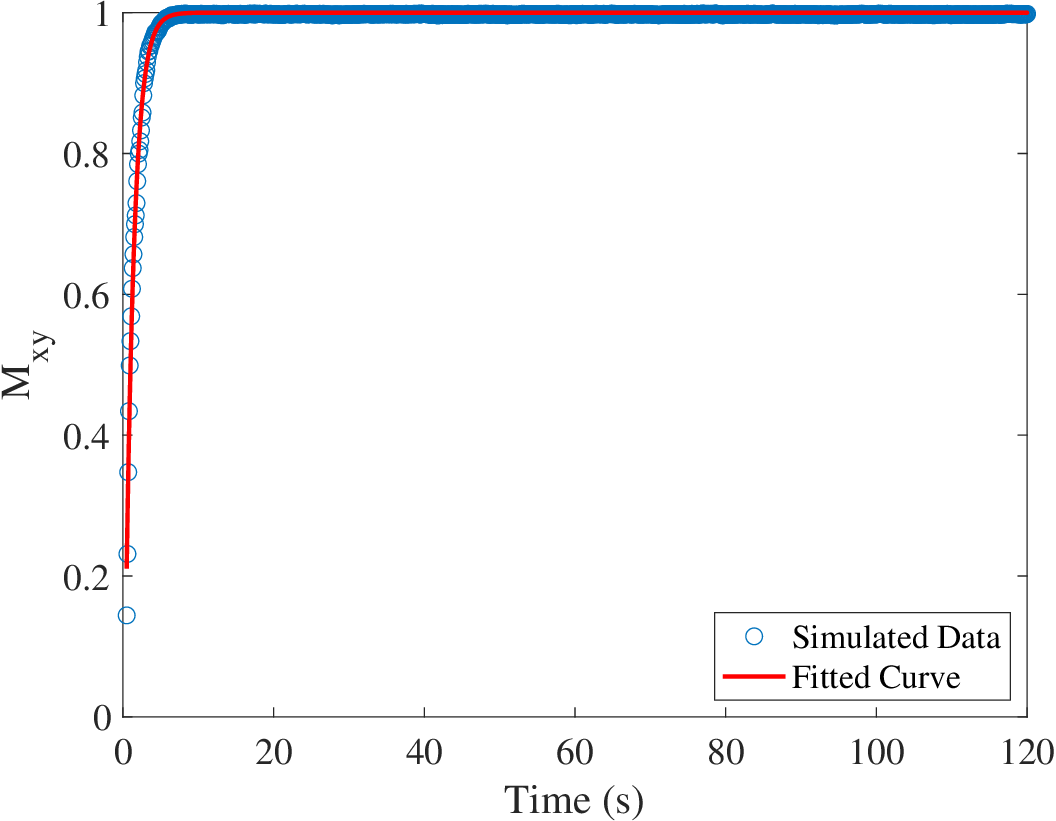} \\
  \end{tabular}
  \caption{Fitted exponential curves (solid lines) and simulated Lacey index data (markers) for the axial direction (left) and cross--sectional direction (right) at a fill level of 40\% and rotation speed of 40\,rpm.}
  \label{fig:M_Fit}
\end{figure}

The fitting was carried out using the Nelder–Mead simplex algorithm \cite{lagarias1998convergence} (implemented via MATLAB’s \texttt{fminsearch}) to minimize the sum of squared residuals between the simulated Lacey index and the fitted curve. Figure~\ref{fig:M_Fit} shows the simulated Lacey index data along with the fitted exponential curves for both axial and cross--sectional directions. The good agreement between simulation and fit supports the use of $t_\mathrm {m}$ as a representative mixing timescale.

The uncertainties in $t_\mathrm{m}$ and $t_0$ were obtained from the covariance matrix of the nonlinear least-squares fit, following the standard procedure described in Reference~\cite{press2007numerical}. The error bars represent one standard deviation.

The fitted $t_\mathrm {m}$ values are used as a quantitative measure of mixing efficiency and will be analyzed in Section~\ref{Result-Effect of operating parameters} to examine the effects of fill level and rotation speed.

\subsubsection{Cycle time}

\begin{figure}[H]
    \centering
    \includegraphics[width=0.6\textwidth]{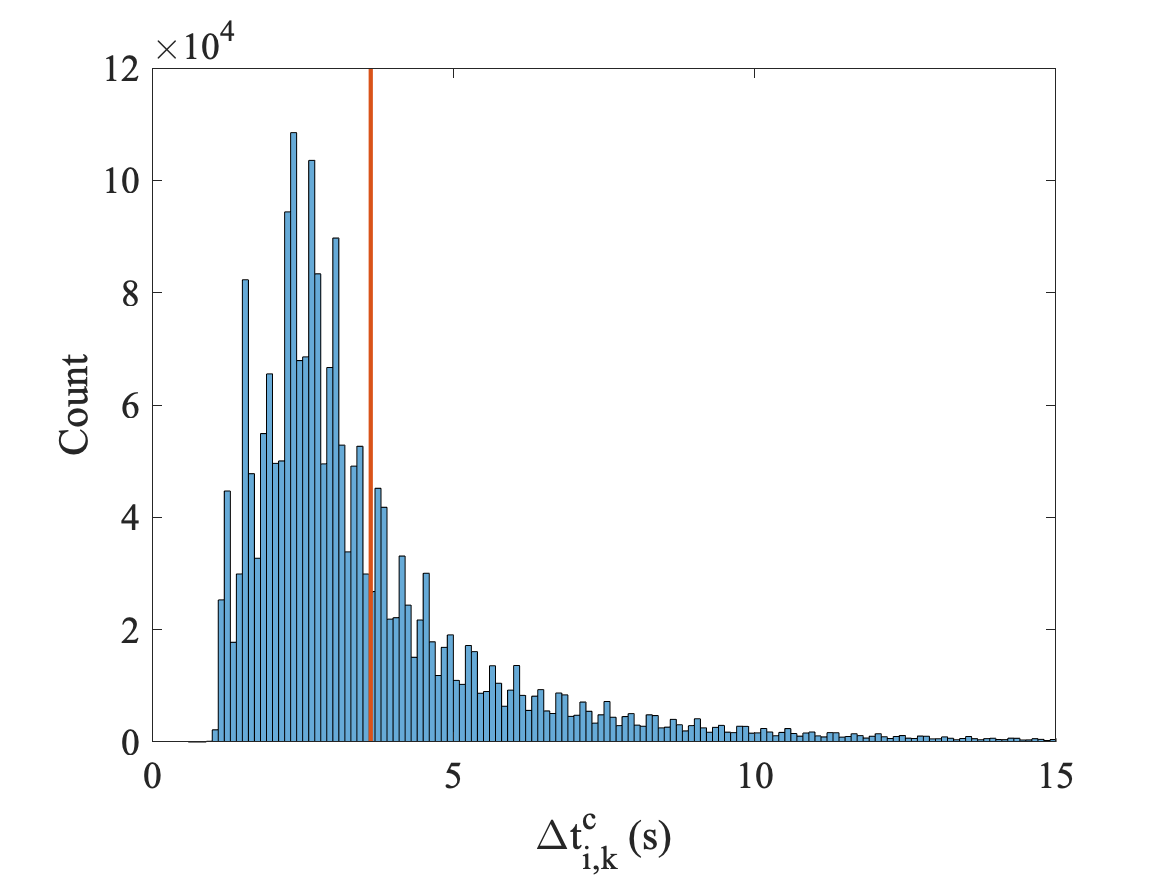}
    \caption{Cycle time distribution (blue) and mean value (red) for the HSBR at a fill level of 40\% and rotation speed of 40\,rpm.}
    \label{fig:CycleTimeDistribution}
\end{figure}

The mean cycle time was first determined directly from the particle trajectories at a fill level of 40\% and rotation speed of 40\,rpm, following the procedure described in Section~\ref{sec:methods}. For each particle, the times $t^{\mathrm{c}}_{i,k}$ associated with the completion of successive revolutions were identified, and the corresponding cycle times $\Delta t^{\mathrm{c}}_{i,k}$ were computed as the difference between consecutive entries, according to Equation~\ref{eq:MeanCycleTime}. The distribution of all valid cycle times collected across all particles is shown in Figure~\ref{fig:CycleTimeDistribution}. The resulting mean cycle time is $\overline{\Delta t^{\mathrm{c}}} = 3.63\,\mathrm{s}$, which represents the average time required for particles to complete one full revolution around the shaft.

To verify this trajectory-based measurement, a continuum coarse--graining (CG) analysis was performed using the framework detailed in Section~\ref{sec:methods}. DEM particle data from the first 120\,s of the simulation were coarse--grained on a $50\times 50$ grid in the xy--plane using a Gaussian kernel of width equal to one mean particle radius. This provided spatially resolved fields for the solid volume fraction $\phi(x,y)$ and velocity components $v_x(x,y)$ and $v_y(x,y)$, from which the angular velocity field $\omega(x,y)$ was computed.

\begin{figure}[H]
    \centering
    \includegraphics[width=0.6\textwidth]{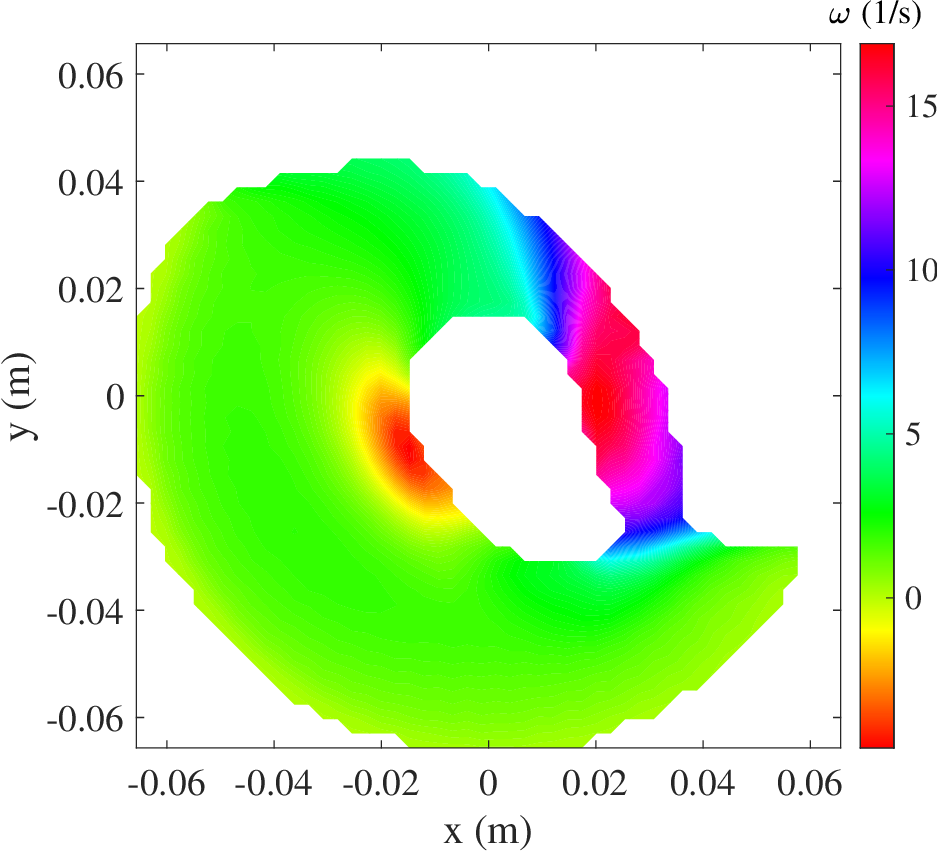}
    \caption{Coarse--grained angular velocity field $\omega(x,y)$ obtained using MercuryCG. The color map represents the local angular velocity magnitude for the HSBR operated at a fill level of 40\% and rotation speed of 40\,rpm.}
    \label{fig:CycleTimeValidation}
\end{figure}

The resulting angular velocity field is shown in Figure~\ref{fig:CycleTimeValidation}. Using the mass--weighted averaging procedure introduced in Section~\ref{sec:methods}, the coarse--grained analysis produced a mean cycle time of $\overline{\Delta t^{\mathrm{c}}}_{\mathrm{CG}} = 3.22\,\mathrm{s}$. This result differs from the trajectory-based value by approximately 11\%, indicating good agreement between the two independent methods. This close correspondence confirms that the DEM-based definition of cycle time reliably captures the angular circulation dynamics within the HSBR.

The mean cycle time will be further analyzed in Section~\ref{Result-Effect of operating parameters} to assess how fill level and rotation speed influence the particle angular motion and circulation dynamics.

\subsubsection{Axial dispersion coefficient}

Using the time--based (Einstein) formulation in Eq.~\ref{eq:EinsteinDispersion_final} with a sampling interval of $\Delta t = 50$\,s, the axial dispersion coefficient for the HSBR operating at a 40\% fill level and 40\,rpm rotation speed was found to be $D^{z} = 3.6\times10^{-5}\,\mathrm{m^2/s}$.

The cycle--based formulation (Eq.~\ref{eq:Dz_cycle}) provided an independent estimate. For the same operating condition, this method yielded an axial dispersion of $D^{z}_{\mathrm{c}} = 1.18\times10^{-4}\,\mathrm{m^2/cycle}$. Using the previously determined mean cycle time of $\overline{\Delta t^{\mathrm{c}}} = 3.63\,\mathrm{s}$ to convert this value to SI units resulted in an axial dispersion coefficient of $3.25\times10^{-5}\,\mathrm{m^2/s}$, which is in close agreement with the time--based estimate.

\begin{figure}[H]
    \centering
    \includegraphics[width=0.6\textwidth]{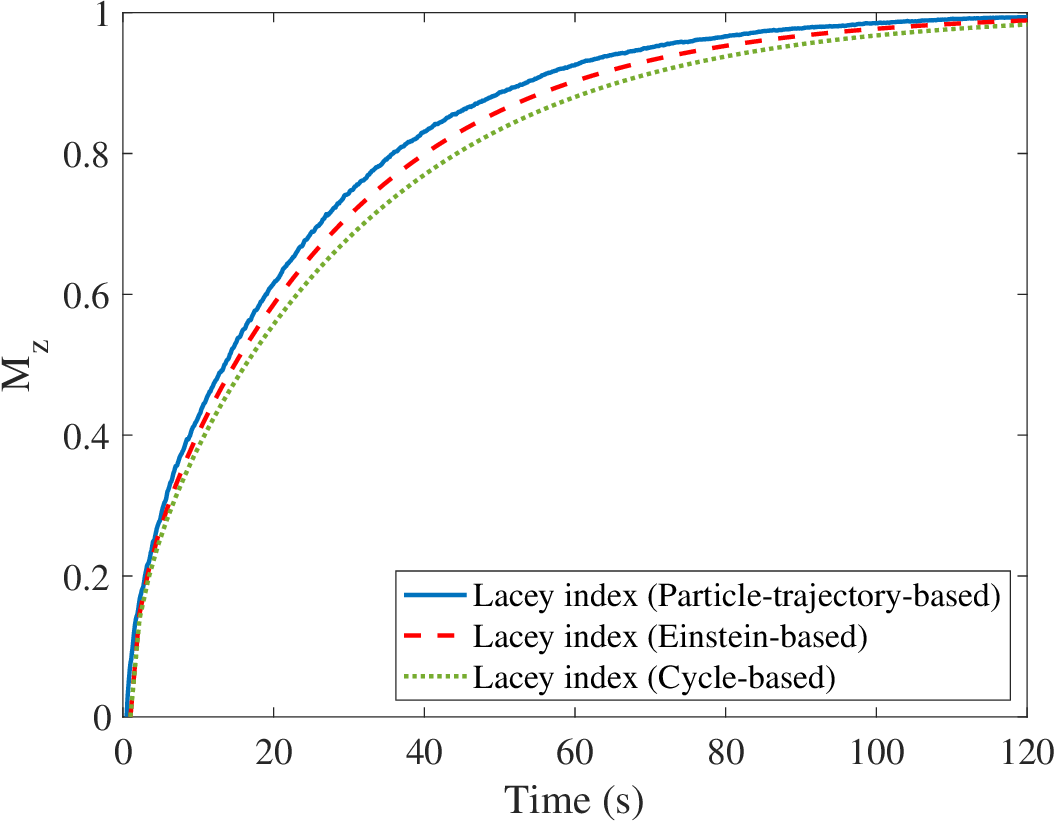}
    \caption{Comparison of the axial Lacey index computed from DEM particle positions with the axial Lacey indices predicted by the diffusion model using the time--based (Einstein) and cycle--based dispersion coefficients.}
    \label{fig:AxialDispersionValidation}
\end{figure}

To further verify these results, the diffusion-based Lacey index comparison introduced in Section~\ref{sec:methods} was applied. The axial diffusion equation was solved using $D^{z}$, and the predicted axial Lacey index $M_z(t)$ was compared with the DEM-based Lacey index. As shown in Figure~\ref{fig:AxialDispersionValidation}, the two curves nearly overlap. A similar comparison using $D^{z}_{\mathrm{c}}$ also produced good agreement.

Finally, by adjusting $D^{z}$ in the diffusion model to obtain the best overlap with the DEM-based Lacey index, a fitted value of $D^{z}_{\mathrm{DEM}} = 3.95\times10^{-5}\,\mathrm{m^2/s}$ was extracted.

Overall, the three independently obtained dispersion coefficients ($3.6\times10^{-5}$, $3.25\times10^{-5}$, and $3.95\times10^{-5}$\,m$^2$/s) are mutually consistent and lie within about 22\% of each other. This consistency demonstrates that all three approaches capture the same underlying axial spreading mechanism in the HSBR and verifies the robustness of the time--based dispersion measurements.

In the following section (Section~\ref{Result-Effect of operating parameters}), the effect of operating conditions on axial dispersion is examined. In particular, the time--based axial dispersion coefficient $D^{z}$ is analysed as a function of fill level and impeller rotation speed to determine how these parameters influence particle spreading along the reactor.

\subsection{Effect of operating parameters (fill level and speed)}\label{Result-Effect of operating parameters}

In this section, we studied the effect of operating parameters (fill level and rotation speed) on the mixing efficiency and flow dynamics of dry polypropylene powder in angular, radial, and axial directions.

\subsubsection{Mixing time}

First, we studied the effect of rotation speed and fill level on the parameter $t_ \mathrm {m}^\mathrm {z}$, which represents the time required to achieve uniform mixing for axial (z--direction) mixing. It was observed that $t_ \mathrm {m}^\mathrm {z}$ consistently increases with fill level at all rotation speeds, while it decreases with increasing rotation speed across all fill levels.

At lower speeds (10--20 rpm, $Fr = 0.015 - 0.06$), $t_ \mathrm {m}^\mathrm {z}$ is highest for the 70\% fill level, reflecting the slower axial mixing due to increased material resistance and compaction effects. For lower fill levels (40\%, 50\%, and 60\%), axial mixing is more efficient, resulting in lower $t_ \mathrm {m}^\mathrm {z}$ values. The reduced material resistance at these fill levels facilitates axial transport.

As the rotation speed increases (30--60 rpm, $Fr = 0.135 - 0.54$), $t_ \mathrm {m}^\mathrm {z}$ decreases for all fill levels, indicating faster axial mixing. The stronger blade-induced forces enhance axial transport by promoting more uniform redistribution of material along the reactor’s length. For higher fill levels (e.g., 70\%), the compaction effects limit the efficiency of axial transport, resulting in consistently higher $t_ \mathrm {m}^\mathrm {z}$ values even at higher speeds.

These results indicate that axial mixing efficiency depends strongly on both rotation speed and fill level. Higher rotation speeds promote faster mixing along the axial direction, as reflected by decreasing $t_ \mathrm {m}^\mathrm {z}$ values. However, as the fill level increases, axial mixing becomes slower due to greater material resistance and compaction, resulting in higher $t_ \mathrm {m}^\mathrm {z}$ values across all speeds.

\begin{figure} [H]
        \centering
	\includegraphics[width=0.6\textwidth]{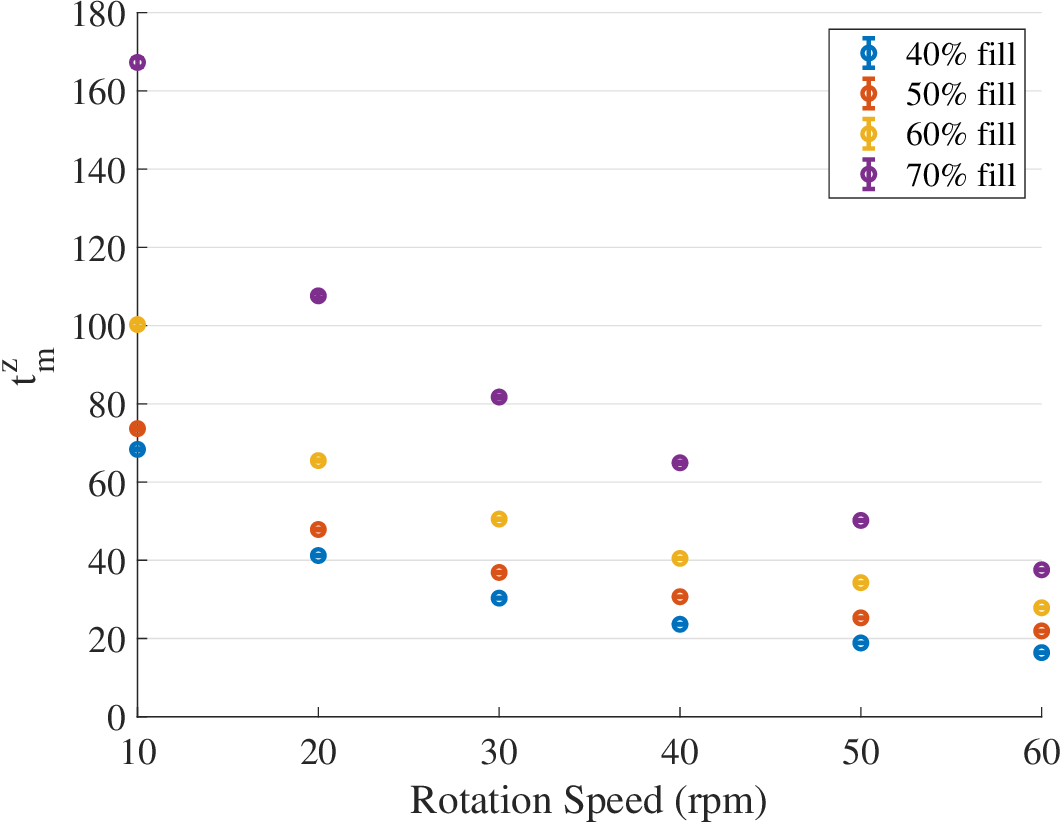}
	\caption{Effect of the fill level and rotation speed on the parameter $t_ \mathrm {m}^\mathrm {z}$ for axial mixing.}
        \label{fig:Parameter_t_m,z}
\end{figure}

Second, we studied the effect of rotation speed and fill level on the parameter $t_ \mathrm {m}^\mathrm {xy}$, for cross--sectional (xy--direction) mixing. We observed that $t_ \mathrm {m}^\mathrm {xy}$ decreases consistently with increasing rotation speed for all fill levels. However, the effect of fill level differs significantly compared to the axial mixing case.

At the lowest speed (10--20 rpm, $Fr = 0.015 - 0.06$), $t_ \mathrm {m}^\mathrm {xy}$ is much higher for the 70\% fill level compared to the other fills, indicating slower mixing. This behavior is attributed to material compaction at higher fill levels, which restricts the ability of the blades to effectively redistribute material radially within the cross--section. For lower fill levels (40\%, 50\%, and 60\%), the parameter $t_ \mathrm {m}^\mathrm {xy}$ is approximately equal, reflecting similar mixing efficiency at these speeds. The reduced material resistance at these fill levels allows the blades to circulate the material more uniformly across the xy--plane.

By increasing speeds (20--30 rpm, $Fr = 0.06 - 0.135$), $t_ \mathrm {m}^\mathrm {xy}$ for the 70\% fill level decreases and approaches the values of the lower fill levels, reflecting improved mixing efficiency as blade-induced forces overcome compaction effects. At these speeds, slight differences emerge among the lower fill levels.

At higher speeds (40--60 rpm, $Fr = 0.24 - 0.54$), $t_ \mathrm {m}^\mathrm {xy}$ for all fill levels becomes nearly equal, suggesting that the effect of fill level diminishes at these speeds. For the 70\% and 40\% fill levels, $t_ \mathrm {m}^\mathrm {xy}$ values are slightly higher compared to 50\% and 60\%. This indicates that the blades effectively redistribute material across the cross--section regardless of fill level at higher speeds, although compaction effects still slightly hinder mixing at the 70\% fill level.

These results show that rotation speed has more effect on cross--sectional mixing compared to fill level, with higher speeds consistently improving mixing efficiency (lower $t_ \mathrm {m}^\mathrm {xy}$ values). The differences in $t_ \mathrm {m}^\mathrm {xy}$ at lower speeds reflect the combined effects of material distribution around the shaft and compaction at higher fill levels, which are minimized as rotation speed increases.

\begin{figure} [H]
        \centering
	\includegraphics[width=0.6\textwidth]{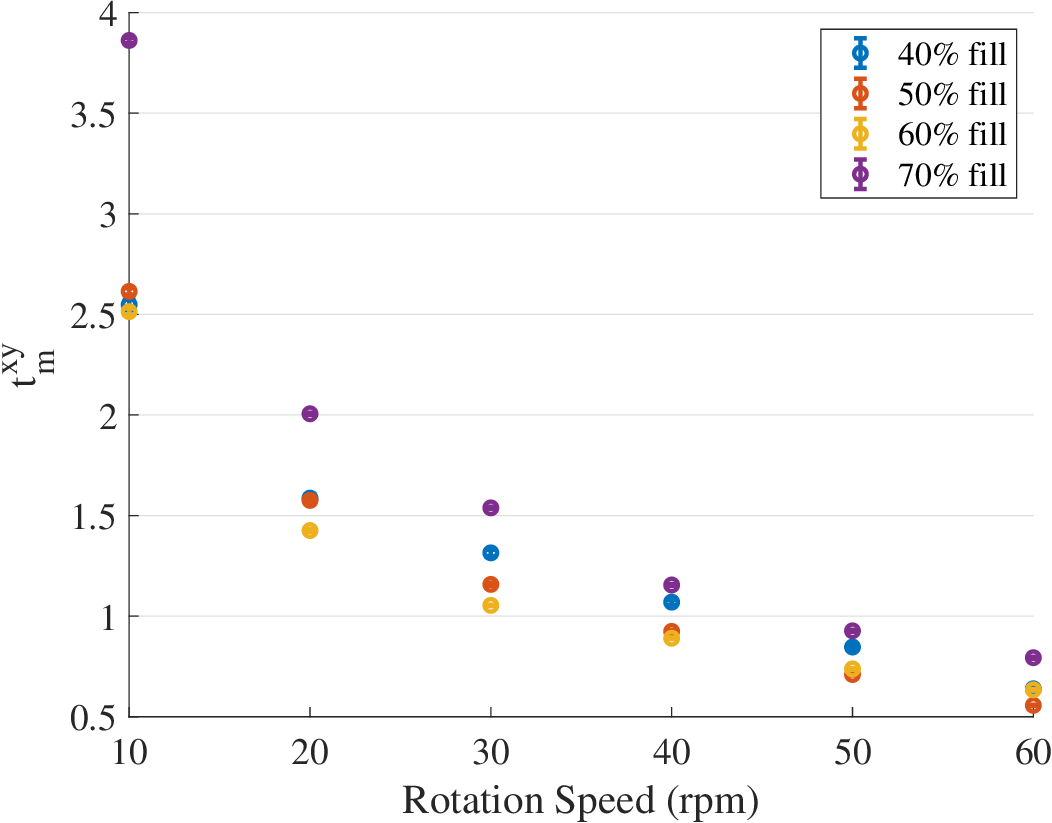}
	\caption{Effect of the fill level and rotation speed on the parameter $t_ \mathrm {m}^\mathrm {xy}$ for cross--sectional mixing.}
        \label{fig:Parameter_t_m,xy}
\end{figure}

\subsubsection{Cycle Time}

\begin{figure}[H]
    \centering
    \renewcommand{\arraystretch}{1.2}
    \begin{tabular}{c@{\hspace{0pt}}c@{\hspace{0pt}}c}
        \includegraphics[width=0.33\textwidth]{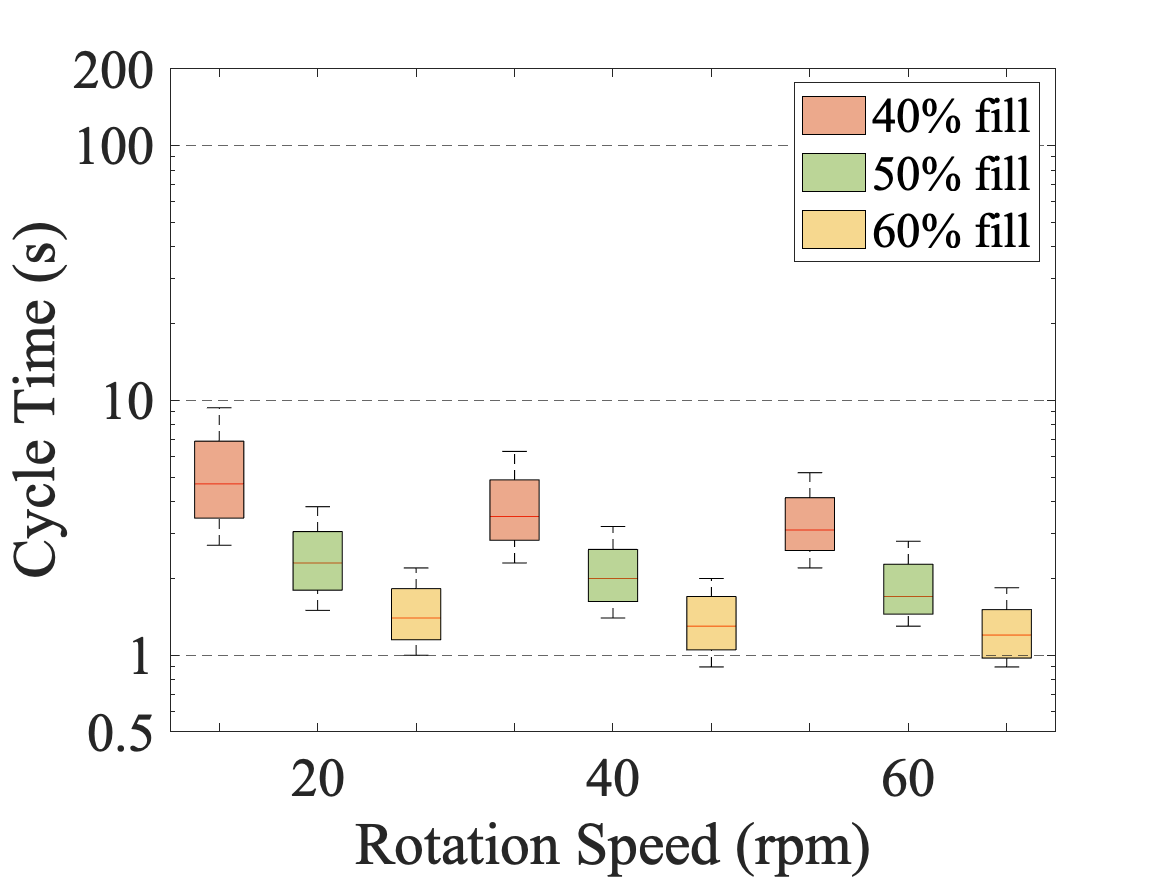} &
        \includegraphics[width=0.33\textwidth]{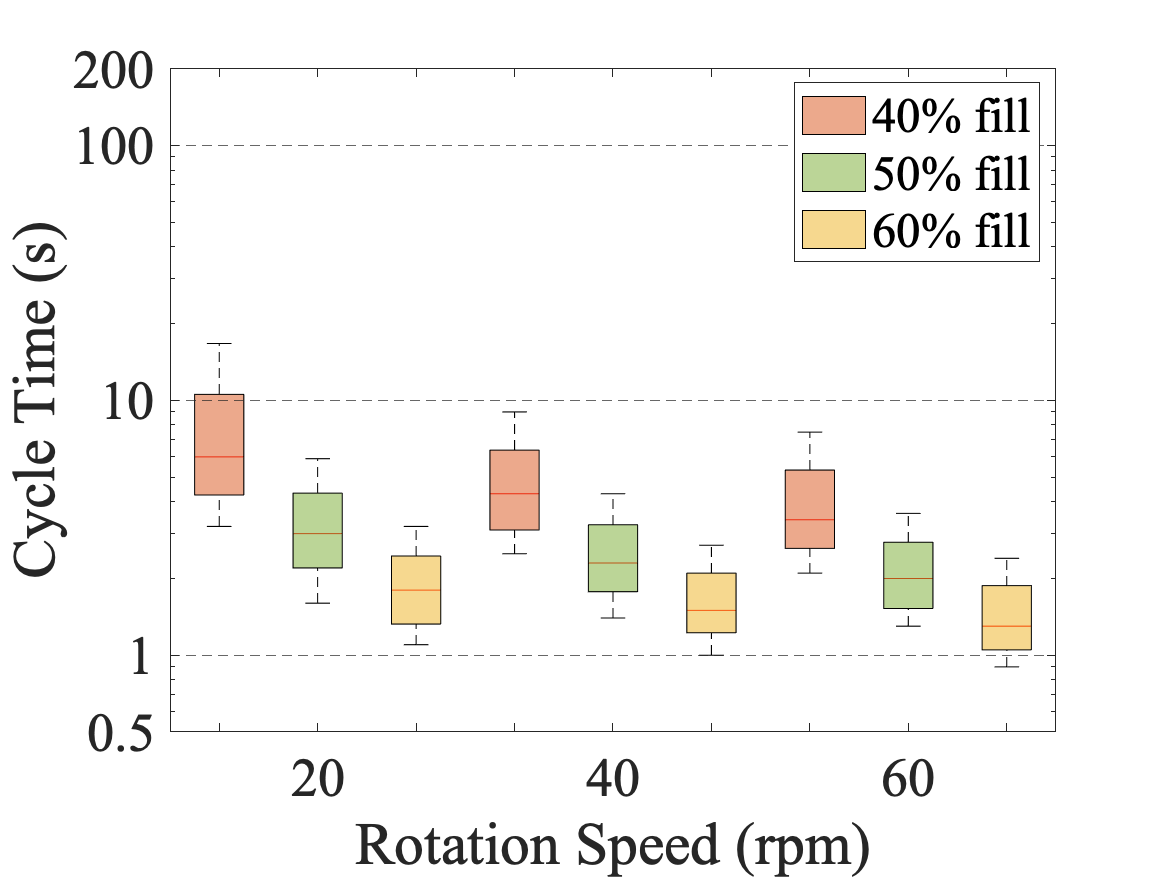} &
        \includegraphics[width=0.33\textwidth]{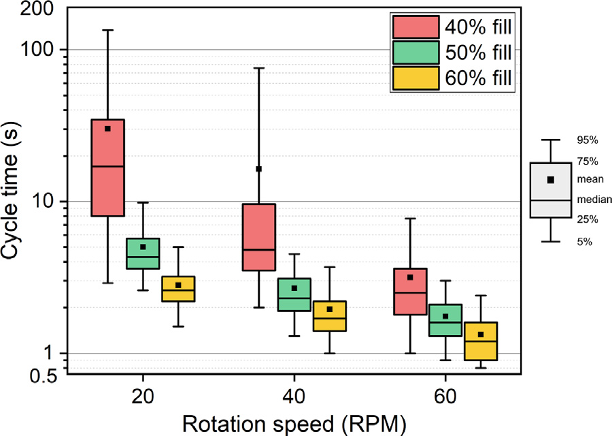} \\
        {\small Scale 5} & {\small Scale 3} & {\small Experiment}
    \end{tabular}
    \caption{Comparison of cycle time obtained from DEM simulations using two particle scale factors with experimental measurements \cite{van2024particle}.}
    \label{fig:ValidationCycleTime}
\end{figure}

An overview of the effect of impeller speed and reactor filling on the cycle time of large particles $(d > d_{v,75})$ (where $d_{v,75}$ is the particle size at 75\% cumulative volume; see Table~\ref{tab:1}) is illustrated in Figure~\ref{fig:ValidationCycleTime}. It can be observed that cycle time decreases as rotation speed and fill level increase. The most considerable point is the influence of the filling level on the angular motion. Low filling (40\%) causes improper angular motion through the reactor, and long particle cycle periods. This causes inhomogeneous heat removal. Therefore, to have a low cycle time, which is demanded for uniform heat removal, we should increase the fill level and impeller speed.

To validate our findings, we compared the simulated cycle times with the experimental measurements reported by van der Sande et al.~\cite{van2024particle}. The simulations reproduce the experimentally observed trends across rotation speeds and fill levels, indicating that the DEM framework captures the essential mechanisms governing particle circulation in HSBRs. Furthermore, reducing the scale factor (i.e., approaching the actual particle size) improves agreement with the experimental results, particularly in terms of magnitude.

\subsubsection{Axial Dispersion Coefficient}

\begin{figure}[H]
    \centering
    \renewcommand{\arraystretch}{1.2}
    \begin{tabular}{c@{\hspace{4pt}}c@{\hspace{4pt}}c}
        \includegraphics[width=0.32\textwidth]{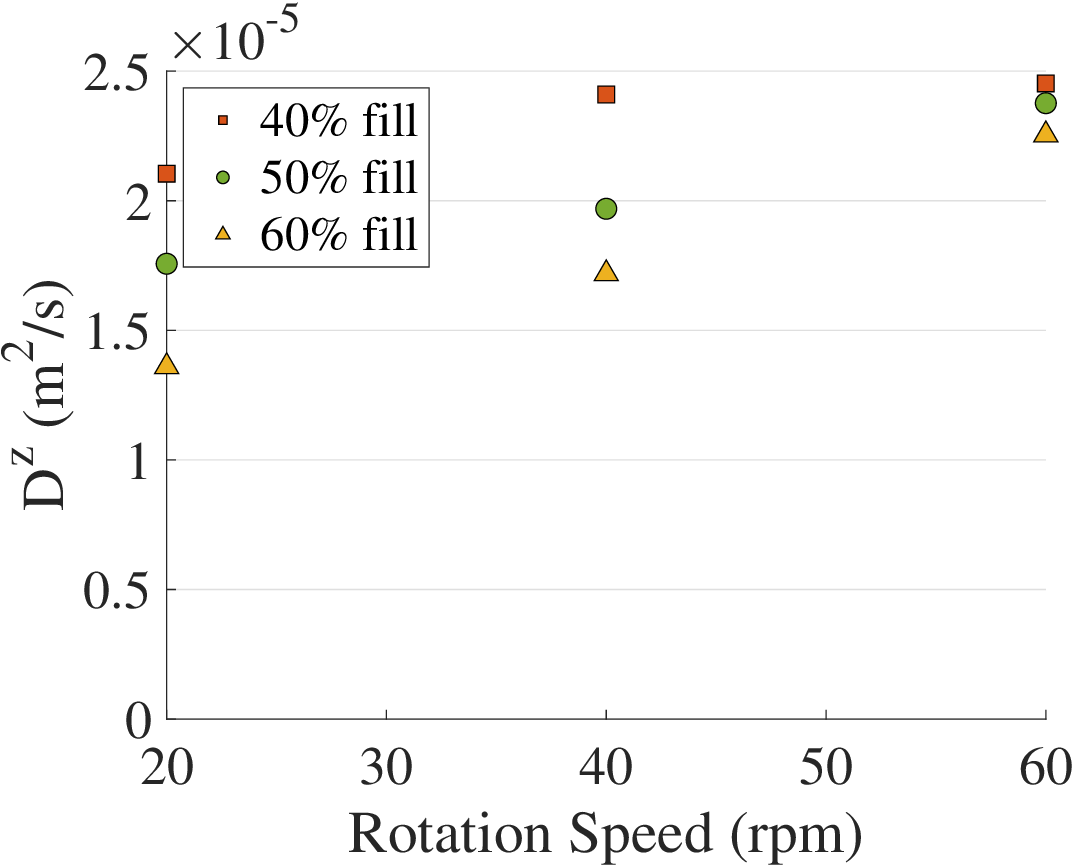} &
        \includegraphics[width=0.32\textwidth]{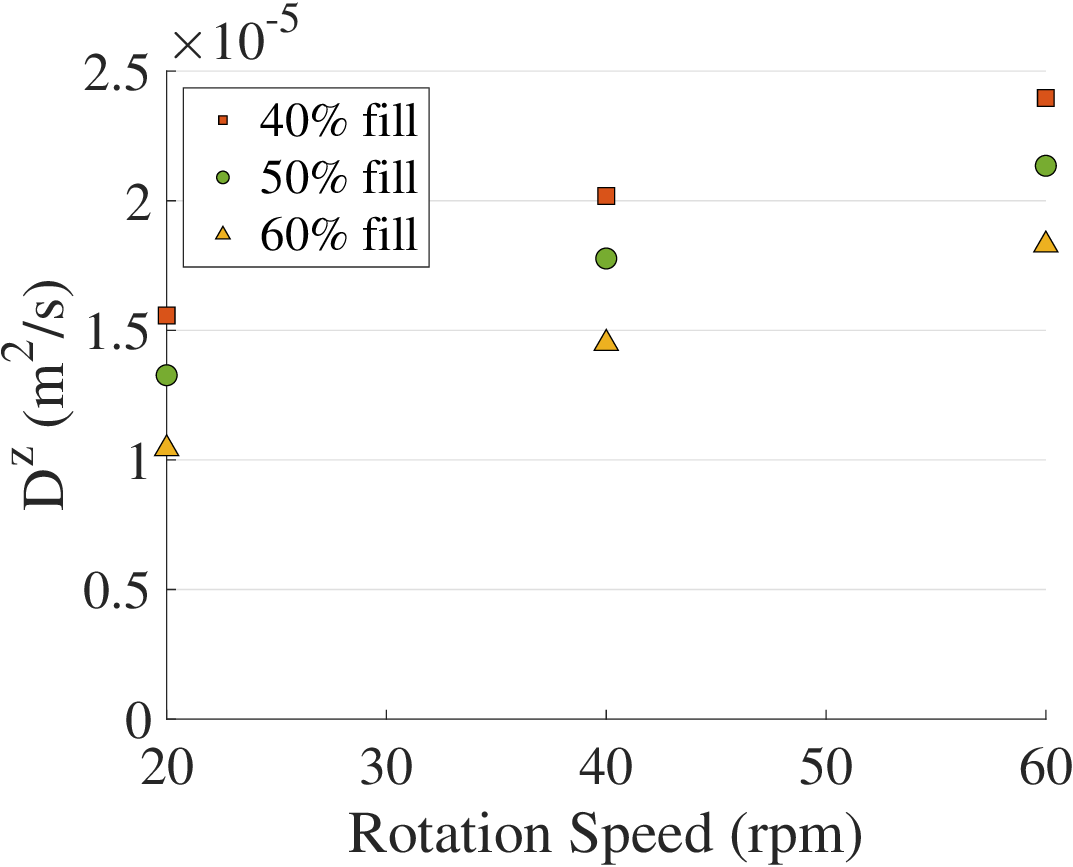} &
        \includegraphics[width=0.32\textwidth]{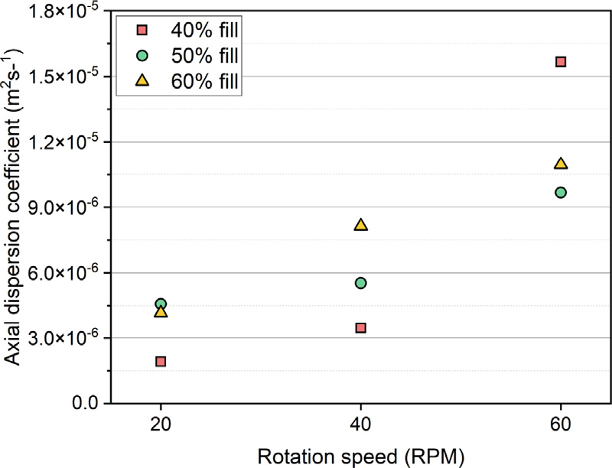} \\
        {\small Scale 5} & {\small Scale 3} & {\small Experiment}
    \end{tabular}
    \caption{Comparison of axial dispersion coefficient obtained from DEM simulations with two scaling factors and experimental measurements \cite{van2024particle}.}
    \label{fig:ValidationAxialDispersion}
\end{figure}

Figure~\ref{fig:ValidationAxialDispersion} shows the effect of reactor filling and rotation speed on the axial dispersion coefficient. It can be observed that the axial dispersion coefficient increases linearly with increasing impeller speed and decreasing fill level. 

The DEM--based dispersion coefficients were compared with the experimental data reported by van der Sande et al.~\cite{van2024particle}, showing consistent agreement in trends across operating conditions. As observed for cycle time, reducing the particle size  improves agreement with the experimental results, particularly in terms of magnitude.

\section{Conclusions and outlook}\label{sec:conclusions}
This study investigated the influence of rotation speed and fill level on key aspects of particle dynamics and mixing efficiency within a lab--scale horizontal stirred bed reactor (HSBR). Using Discrete Element Method (DEM) simulations validated against experimental data by van der Sande et al.~\cite{van2024particle}, we analysed the key mechanisms governing reactor performance under different operating conditions.

Axial mixing was found to depend strongly on both rotation speed and fill level. Higher rotation speeds promote faster mixing along the axial direction. However, as the fill level increases, axial mixing becomes slower due to greater material resistance and compaction. Cross--sectional mixing, on the other hand, was found to be more sensitive to rotation speed than to fill level. Higher speeds consistently enhanced mixing efficiency, while variations at lower speeds highlighted the influence of material distribution around the shaft and compaction at higher fill levels.

Cycle time, the average time required for a particle to complete one revolution around the shaft, decreased with increasing rotation speed and fill level. At lower fill levels, prolonged cycle times led to inefficient heat removal due to insufficient particle circulation. In contrast, higher fill levels led to shorter cycle times, thereby improving heat transfer. The strong qualitative agreement between our simulation results and previously published experimental data confirms the reliability of the DEM study, with quantitative accuracy improving as the particle size scaling factor was reduced.

Axial dispersion coefficients increased with higher rotation speeds but decreased with increasing fill levels. Enhanced particle mobility at higher speeds and greater displacement in less compacted beds contributed to this trend. The alignment of these results with existing experimental findings illustrates the ability of the simulations to capture complex flow dynamics within the HSBR.

Overall, the results reveal a clear operational trade--off. Increasing the rotation speed consistently improves mixing efficiency, reduces the cycle time, and enhances axial dispersion. Increasing the fill level, in contrast, shortens the cycle time and promotes solids circulation, but simultaneously reduces axial mixing efficiency and axial dispersion. These findings highlight the need to balance operating conditions to achieve optimal reactor performance. The agreement between simulations and experimental measurements further supports the validity of the DEM approach. Consequently, the DEM framework developed in this work provides a powerful tool for analysing these interactions and guiding industrial reactor optimization.

Future work should extend this analysis to full--scale HSBRs and diverse particulate systems. Incorporating more realistic particle shapes and further reducing particle size scaling in simulations can improve the accuracy and applicability of the results. Experimental validation involving multi--particle systems would also bridge the gap between simulation and practical implementation, further enhancing the utility of HSBRs in industrial polymerization processes.
\section*{Acknowledgment}

This research is part of the Industrial Dense Granular Flows project, which received funding from the Dutch Research Council (NWO) in the framework of the ENW PPP Fund for the top sectors and from the Ministry of Economic Affairs under the “PPS-Toeslagregeling”.


The authors sincerely thank Professor Stefan Luding and Professor Anthony Thornton for valuable scientific discussions and partial supervision.
 
%

\appendix
\section{Influence of the sampling interval on the axial dispersion coefficient} \label {app:DeltaTStudy}
\begin{figure}[H]
\centering
\includegraphics[width=0.6\textwidth]{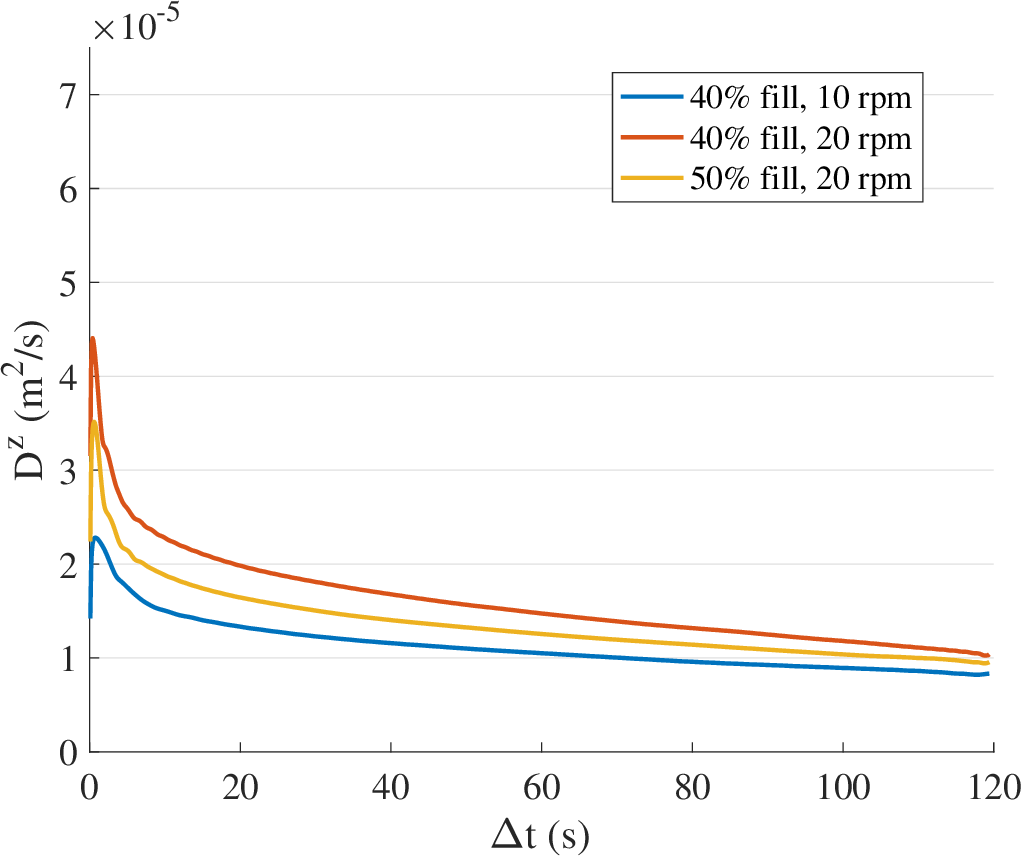}
\caption{Axial dispersion coefficient $D^{z}$ as a function of the sampling interval $\Delta t$ for different HSBR operating conditions (see legend). The analysis is performed for the selected particle size class defined by $d > d_{v,75}$.}
\label{fig:Dz_vs_deltat}
\end{figure}

To determine a suitable sampling interval $\Delta t$ for the time--based (Einstein) axial dispersion coefficient, the dispersion was evaluated over a wide range of $\Delta t$ values.
Figure~\ref{fig:Dz_vs_deltat} shows the resulting dependence of $D^{z}$ on $\Delta t$.

At very small $\Delta t$, the calculated dispersion coefficient is strongly affected by short--time correlations. As $\Delta t$ increases, $D^{z}$ decreases rapidly and then enters a quasi--plateau region in which its dependence on $\Delta t$ becomes much weaker.

\begin{figure}[H]
\centering
\includegraphics[width=0.6\textwidth]{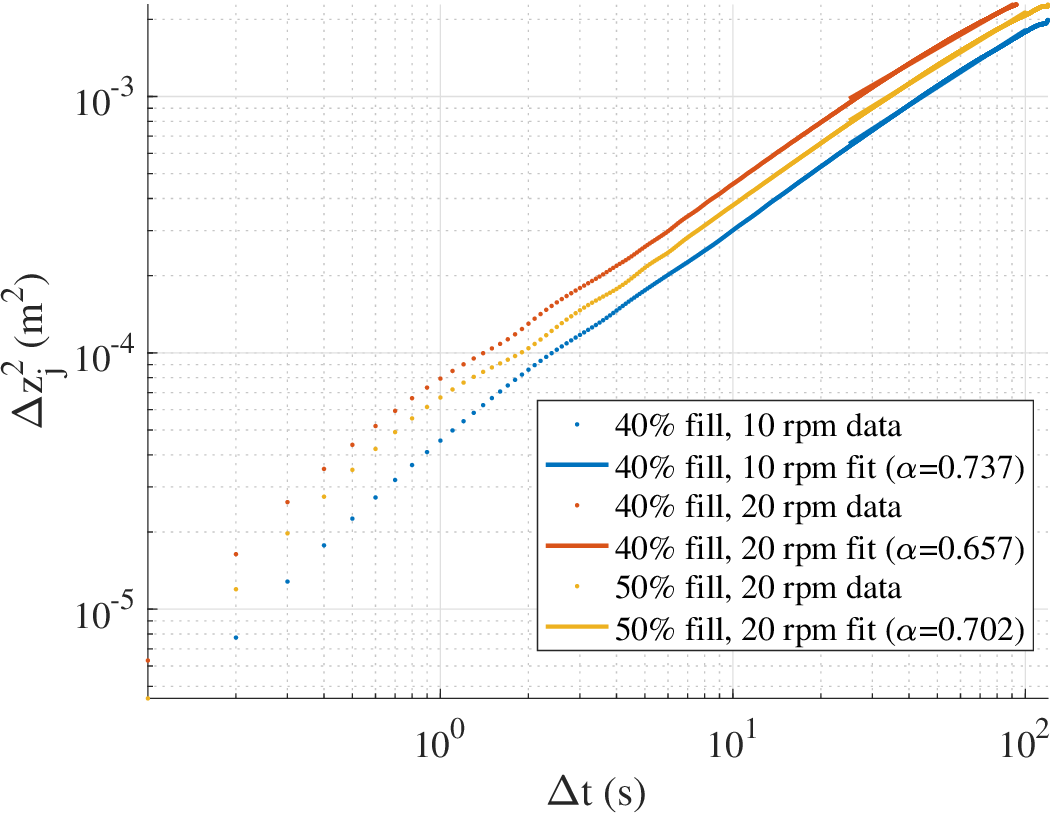}
\caption{Mean squared axial displacement as a function of the sampling interval $\Delta t$ for different HSBR operating conditions (see legend). Symbols denote simulation data and solid lines represent fits in the range $25\,\mathrm{s} < \Delta t < 100\,\mathrm{s}$. The slope of each fitted line corresponds to the exponent $\alpha$, indicating sub--diffusive behaviour. The analysis is performed for the selected particle size class defined by $d > d_{v,75}$.}
\label{fig:MSD_axial}
\end{figure}

To further analyse this behaviour, the mean squared axial displacement (MSD) was evaluated as a function of $\Delta t$, as shown in Figure~\ref{fig:MSD_axial}. The MSD follows a power--law relation

\begin{equation}
\langle (\Delta z)^2 \rangle \propto \Delta t^{\alpha},
\end{equation}

\noindent with $\alpha < 1$, indicating sub--diffusive axial transport. As a consequence, the apparent dispersion coefficient derived from $\langle (\Delta z)^2 \rangle/(2\Delta t)$ does not reach a perfectly constant plateau but decreases slowly with increasing $\Delta t$.

This sub--diffusive behaviour reflects long--time correlations in the axial displacement, which may be promoted by dense frictional contacts and the structured, blade--driven circulation in the HSBR. In addition, the particle upscaling used in the DEM model can reduce collisional diffusion and prolong decorrelation relative to the experimental system.

Based on this analysis, a sampling interval of $\Delta t = 50\,\mathrm{s}$ was selected for the evaluation of the axial dispersion coefficient in this study. This value lies within the quasi--plateau region, avoids short--time correlation effects, and retains sufficient sensitivity to operating conditions.

%

\bibliography{ref}


%
%
\end{document}